\documentclass[a4paper,11pt]{article}
\usepackage[utf8]{inputenc}

\usepackage{amsfonts}
\usepackage{amssymb}
\usepackage{graphicx}
\usepackage{amsmath}
\usepackage{enumerate}
\usepackage{color}
\usepackage{cite}
 \usepackage{multirow}
\usepackage{float}

\RequirePackage{mathrsfs} \RequirePackage[sc]{mathpazo}
\RequirePackage{wasysym} \RequirePackage{setspace}\textheight=650pt
\title{ {\bf Behavior of Quasinormal Modes and high dimension RN-AdS Black Hole phase transition}}
\author{M. Chabab$^{1}$\footnote{mchabab@uca.ma (Corresponding author)}, H. El Moumni$^{1,2}$\footnote{hasanelm@yahoo.fr}, S. Iraoui$^{1}$\footnote{thesamirsim@gmail.com}, K. Masmar$^{1}$\footnote{karima.masmar@edu.uca.ac.ma}\\
	\\ 
	{\small $^{1}$ High Energy Physics and Astrophysics Laboratory, Faculty of Science Semlalia, 
	}\\
         {\small Cadi Ayyad University, 40000 Marrakesh, Morocco.
	}\\
	{\small $^{2}$ LMTI, Physics Departement, Faculty of Sciences,  Ibn Zohr University, Agadir, Morocco. }
}
\date{}

\begin{document}
 \maketitle

\abstract{In this work we use the quasinormal frequencies  of a massless scalar perturbation to probe the phase transition of the high dimension charged-AdS black hole.
The signature of the critical behavior of this black hole solution is detected in the isobaric as well as in isothermal process. This paper is a natural generalization of \cite{base} to higher dimensional spacetime. More precisely our study shows a clear signal for any dimension $d$ in the isobaric process. As to the isothermal case, we find out that this signature can be affected by other parameters like the pressure and the horizon radius.
We conclude that the quasinormal modes  can be an efficient tool to investigate the first order phase transition, but fail to disclose the signature of the second order phase transition.}

\section{Introduction}
Since the original works of Vishveswara \cite{vis} on quasinormal 
modes (QNMs), considered as solutions of a perturbed wave equation  around black hole configuration. an increasing interest has been devoted to their study. Indeed the characteristic oscillations of the black holes play a prominent role in the dynamics of astrophysical black holes in the framework of general relativity  \cite{mx23,mx25}. 
Last February, LIGO (Laser Interferometer Gravitational-Wave Observatory) announced for the first time the detection of gravitational waves from a cosmic event  generated more than one billion year ago by the merger of two orbiting black holes  \cite{mx1}. The observation of these waves is an important landmark  in the gravitational waves astronomy. The measure of the QNMs frequencies by various detectors such as LIGO \cite{mx26}, VIRGO \cite{mx27}, KAGRA \cite{mx28,mx28b}, can shed light on  the main feature of the black holes such the mass and the angular momentum. \\
The properties of quasinormal modes have been tested in the context of the AdS/CFT correspondence  
 \cite{cliff24}, the investigation of the stability of AdS black holes becomes more appealing.
The quasinormal frequencies of AdS black holes have direct interpretation in terms of the dual conformal field theory CFT, for detail we can refer to 
 \cite{mx29,mx30,mx31,mx30b,mx30b2,mx31b,mx28b2}. The QNMs can be used as a powerful tool to detect the extra dimensions of spacetime,
in other words the brane-world scenarios assume the existence of extra dimensions, so that multidimensional black holes can be formed in a laboratory \cite{lab1,lab2,lab3,lab4}.

 Recently, many efforts have been devoted to study the
 thermodynamical properties of  black holes, in connection with higher dimensional supergravity models \cite{01,02}. These
 properties  have been extensively studied via different methods including numerical computation  using various codes \cite{04}. In fact, several  models based on  mathematical methods have been explored to study critical behaviors  of  black holes with different geometrical configurations in arbitrary dimensions. A particular emphasis has been put on  AdS black  holes \cite{Kastor:2009wy,30,a1,4,5,50,6}. More precisely, a nice interplay between the behaviors of the RN-AdS black hole systems and the Van der Waals fluids has been discussed at length in \cite{KM,chin1,our,our1,our2}.
 
 In this context, several landmarks of statistical liquid-gas systems, such as the $P-V$ criticality, the Gibbs free energy, the first order phase transition and the behavior near the critical points  have been derived.
  Also, in \cite{base} the authors established a link between the Reissner-Nordström AdS black hole critical behavior and the quasinormal modes in $d=4$, showing that QNM can be a dynamic probe of the thermodynamic phase transition. More precisely they find a drastic change in the slopes of the quasinormal frequencies in the small (SBH) and large (LBH) black holes near the critical point where the Van der Waals like thermodynamic phase transition occurs. Also in \cite{zzz1}, the authors studied  the  quasinormal modes and $P-V$ criticality for scalar perturbations in a class of dRGT massive gravity around Black Holes.

Motivated by these results  we find crucial and well justified to generalize these studies to high dimensional spacetime, since higher dimension RN-AdS black hole also presents a Van der Waals like phase transition \cite{KM,our}.

Our paper is organized as follows. We  briefly review the critical behavior of the charged-AdS black hole in arbitrary  dimensional spacetime in section \ref{behavor}. In section \ref{master} we develop the master equation for the massless scalar field perturbation in  high dimensional RN-AdS spacetime. Section \ref{isobaric} and \ref{isothermal} are devoted to the calculation of the quasinormal modes of scalar perturbation around the black holes and to show how their frequencies encode the signature of the first order phase transition. Also we will see that the  QNMs can not  reveal the second order phase transition. Finally  in the last section we summarize our results and draw the conclusion. 
\section{Critical behavior of RN-AdS black holes in higher dimension: A review \label{behavor}}
The Einstein-Maxwell-anti-de Sitter action in higher dimensions $d$  may be written as,
\begin{equation}\label{actem}
I_{EM}=-\frac{1}{16\pi}\int_M d^dx\sqrt{-g}\Bigl(R-F^2+\frac{(d-1)(d-2)}{l^2}\Bigr)\,.
\end{equation} 

The solution of the action \eqref{actem} is the solution for a spherical d-dimensional charged AdS black hole  with negative cosmological constant where $d\geq 4$. It can be written in the static form
\begin{eqnarray}\label{HDRN}
ds^2 &=& -f dt^2 + \frac{dr^2}{f} + r^{2} d\omega_{d-2}^2\,,\nonumber\\
F&=&dA\,,\quad A=-\sqrt{\frac{d-2}{2(d-3)}}\frac{q}{r^{d-3}}dt\,.
\end{eqnarray}

Here, the function $f$ is given by 
\begin{equation}\label{metfunction}
f = 1 - \frac{m}{r^{d-3}} + \frac{q^2}{r^{2(d-3)}} + \frac{r^2}{l^2}\,,
\end{equation}
and $d\omega_d^2$ is the metric of a d-dimensional unit sphere.

The parameters $m$ and $q$ are related to the ADM mass $M$ (in our set up, they are associated with the enthalpy of the system as we shall see later) and the charge $Q$ of the black hole as \cite{4},
\begin{eqnarray}
M&=&\frac{d-2}{16 \pi} \omega_{d-2} m\,,\nonumber\\
Q&=&\frac{\sqrt{2(d-2)(d-3)}}{8\pi}\,\omega_{d-2}\,q\,,
\end{eqnarray} 
with $\omega_d$ being the volume of the unit $d$-sphere, 
\begin{equation}
\omega_d=\frac{2\pi^{\frac{d+1}{2}}}{\Gamma\left(\frac{d+1}{2}\right)}\,.
\end{equation}
It is worth to notice that the cosmological constant $\Lambda = - \frac{(d-1)(d-2)}{2 l^2}$ is regarded as a variable and also identified with the thermodynamic pressure $P$,
\begin{equation}\label{PLambda}
P = - \frac{1}{8 \pi} \Lambda=\frac{(d-1)(d-2)}{16 \pi l^2}\,,  
\end{equation}
in  the geometric  units $G_{d}=\hbar=c=k_{B}=1$. The corresponding conjugate quantity, namely the thermodynamic volume, is given by \cite{50}
\begin{equation}\label{volrp}
V  = \frac{\omega_{d-2} {r_H}^{d-1}}{d-1}\,,
\end{equation}
while the black hole temperature reads as
\begin{equation}\label{T4}
T = \frac{f'(r_H)}{4 \pi}
= \frac{d-3}{4 \pi r_{H}} \Bigl(1 - \frac{q^2}{r_H^{2(d-3)}} + \frac{d-1}{d-3}\frac{r_H^2}{l^2} \Bigr)\,,
\end{equation}
where  the position of the black hole event horizon $r_H$, determined by solving the condition $f(r=r_H)=0$ and choose the largest real positive root. The electric potential $\Phi$ measured at infinity with respect to the horizon while the black hole entropy $S$, was determined from the Bekenstein-Hawking formula $(S= \frac{A_{d-2}}{4})$. They are given by
\begin{eqnarray}\label{S4} 
\Phi&=&\sqrt{\frac{d-2}{2(d-3)}}\frac{q}{r_H^{d-3}}\,,\\
S &=& \frac{\omega_{d-2} r_H^{d-2}}{4}\,.
\end{eqnarray}
All these quantities satisfy the Smarr formula
\begin{equation}
M=\frac{d-2}{d-3}TS+\Phi Q-\frac{2}{d-3} VP\,,
\end{equation}  
where the black hole mass M  is identified with the enthalpy rather than the internal energy of the gravitational system \cite{a1}, so the first law of black hole thermodynamics reads, 
\begin{equation}
dM=TdS+\Phi dQ+VdP\,.
\end{equation}

From Eq. (\ref{T4}), one can derive the following equation of states $P=P(r_H,T)$ for a charged AdS black hole in arbitrary dimension $d$,
\begin{equation}
 P = \frac{T(d-2)}{4 r_{H}} - \frac{(d-3)(d-2)}{ 16 \pi r_H^2}
 + \frac{q^2 (d-3) (d-2)}{ 16 \pi r_{H}^{2(d-2)}}\,.
\end{equation}
As usual, the critical points occur when P has an inflection point,
\begin{equation}\label{critical}
 \left.\frac{\partial P}{\partial r_H}\right|_{T=T_c,r_H=r_c}= \left.\frac{\partial^2 P}{\partial r_H^2}\right|_{T=T_c,r_H=r_c}=0\,,
\end{equation}
loading to,
\begin{eqnarray}\label{pointcrit}
v_c &=& \frac{4 \left((d-2) (2 d-5) q^2\right)^{\frac{1}{2 (d-3)}}}{d-2}\,, \nonumber\\
T_c &=& \frac{(d-3)^2 \left((d-2) (2 d-5) q^2\right)^{\frac{1}{6-2 d}}}{\pi  (2 d-5)}\,,\nonumber\\
P_c &=& \frac{(d-3)^2 \left((d-2) (2 d-5) q^2\right)^{\frac{1}{3-d}}}{16 \pi }\,.
\end{eqnarray}
As to the Gibbs free energy $G=M-TS$,  for fixed charge, it reads as \cite{4}
\begin{equation}
\label{GibbsQ}
G=G(P,T)
=\! \frac{{\omega}_{d-2} }{16 \pi} \biggl(\!r_H^{d-3} - \frac{16\pi Pr_H^{d-1}}{(d\!-\!1)(d\!-\!2)}
+ \frac{(2d\!-\!5)q^2}{r_H^{d-3}}\biggl).\ \ 
\end{equation}

After having briefly introduced the main thermodynamical quantities and  related  phase transition, let's  focus our attention in the next section on the derivation of the  quasinormal frequencies of a   scalar perturbation around a charged AdS black hole in high dimension spacetime.
\section{Master equation in high dimension RN-AdS black hole spacetime \label{master}}

In the sequal, we study the evolution of a massless scalar field
	 in the background of high dimensional charged AdS black hole. The radial part of the
	field, $\Psi(r,t)=\psi(r) e^{-i \omega
		t}$, obeys the Klein-Gordon  wave equation,

\begin{equation}\label{KG}
	\frac{1}{\sqrt{-g}}\partial_{\mu}\left(\sqrt{-g} g^{\mu\nu}\partial_{\nu}\Psi(t,r)\right)=0,
\end{equation} 
where $g_{\mu\nu}$ are the covariant metric components of the RN-AdS  spacetime, $g^{\mu\nu}$ are the contravariant metric components and $g$ the metric determinant. The radial functions $\psi(r)$ obey the following differential equation

\begin{eqnarray}\label{psieqn}
\psi''(r)+\left[\frac{f'(r)}{f(r)}+\frac{d-2}{r}\right] \psi'(r)+\frac{\omega^2 \psi(r)}{f(r)^2}=0,
\end{eqnarray}
where $\omega$ are complex numbers $\omega=\omega_{r}+i \omega_{im}$,  corresponding  to the  QNM  frequencies of the oscillations describing the perturbation.  Notice that the imaginary part $\omega_{im}$ must be negative since  the QNMs  decay in time.

Defining $\psi(r)$ as $\psi(r) \exp[-i\int \frac{\omega}{f(r)}dr]$, then in the vicinity of the black hole horizon $r_H$, the exponential simplifies to, 
\begin{eqnarray}\label{ingoing}
\exp[-i\int
\frac{\omega}{f(r)}dr]\sim(r-r_H)^{-i \frac{\omega}{4\pi T}},
\end{eqnarray}

 and the incoming waves behaves as,
$\psi(r)\sim(r-r_H)^{-i \frac{\omega}{4\pi T}}$, which means that $\psi(r)$ is set to unity at the horizon \cite{base}. Then 
 we can reformulate
Eq. (\ref{psieqn}) as
\begin{eqnarray}\label{psieqn0}
\psi''(r)+\psi'(r) \left[\frac{f'(r)}{f(r)}-\frac{2 i \omega}{f(r)}+\frac{d-2}{r}\right]- \psi(r)\frac{(d-2) i \omega}{r f(r)}=0,
\end{eqnarray}
At this stage, for computation reasons,  we use the following coordinate transformation $u=\left(\frac{r_H}{r}\right)^{d-3}$, so the AdS boundary is located at $u=0$ while the horizon is located at $u=1$. In this case, Eq.  \eqref{psieqn0} becomes,
\begin{eqnarray}\label{psieqn01}
\psi''(u)+\left[\frac{f'(u)}{f(u)}+\frac{2i\omega r_{H}}{(d-3)u^{\frac{d-2}{d-3}}f(u)}\right]\psi'(u)-\frac{(d-2)i\omega r_{H}}{(d-3)^{2}u^{\frac{2d-5}{d-3}} f(u)}\psi(u)=0,
\end{eqnarray}

with $\psi(u)=0$ at the AdS limit $u\rightarrow 0$.  Thanks to the boundary conditions, we can solve numerically Eq. (\ref{psieqn01}) and find out the frequencies of the quasinormal modes  via the shooting method\footnote{Shooting method has been applied in many cases for finding QN frequencies, see for example~\cite{Janiszewski,Mahapatra}}.

The main purpose of the present paper is to show that the quasinormal modes dynamical behavior in the massless scalar perturbation encode 
the signature of thermodynamical first order phase transition of high dimensions charged AdS black holes, 
The isobaric process and the isothermal are  used to perform this analysis in the context of  the van der Waals phase transition picture.


\section{QNM behaviors in the isobaric phase transition\label{isobaric}}

Since the pressure $P$ (or cosmological constant $\Lambda$) is fixed in an  isobaric phase transition, then the horizon radius $r_H$ is the only variable in the system. Our subsequent analysis will take into account the spacetime dimension.

\begin{figure}[!ht]
	\hspace*{0.8cm}	\begin{minipage}{0.88\linewidth}
			\begin{tabbing}
				\hspace{7.5cm}\=\kill

				\includegraphics[scale=.45]{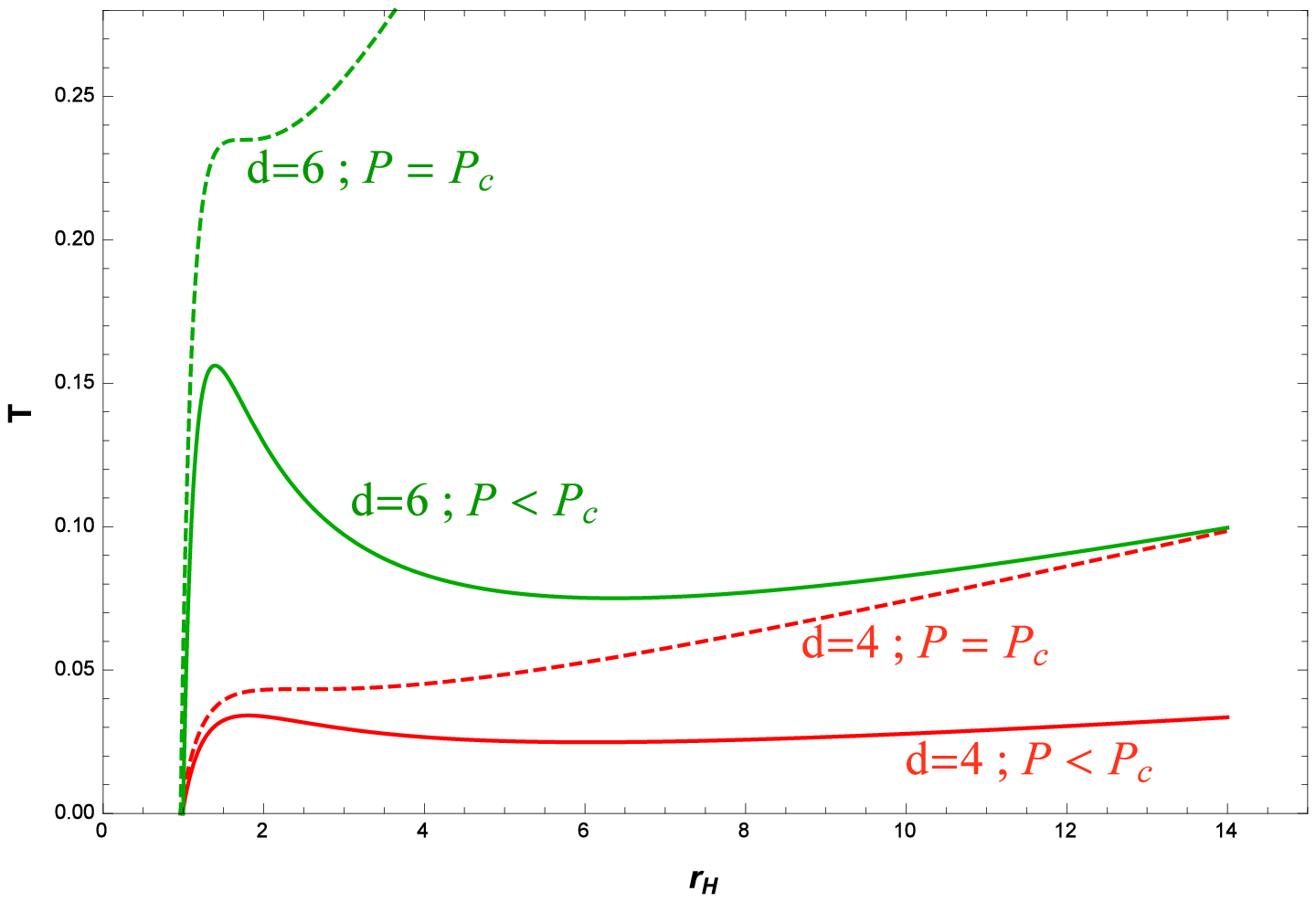}\> \includegraphics[scale=.45]{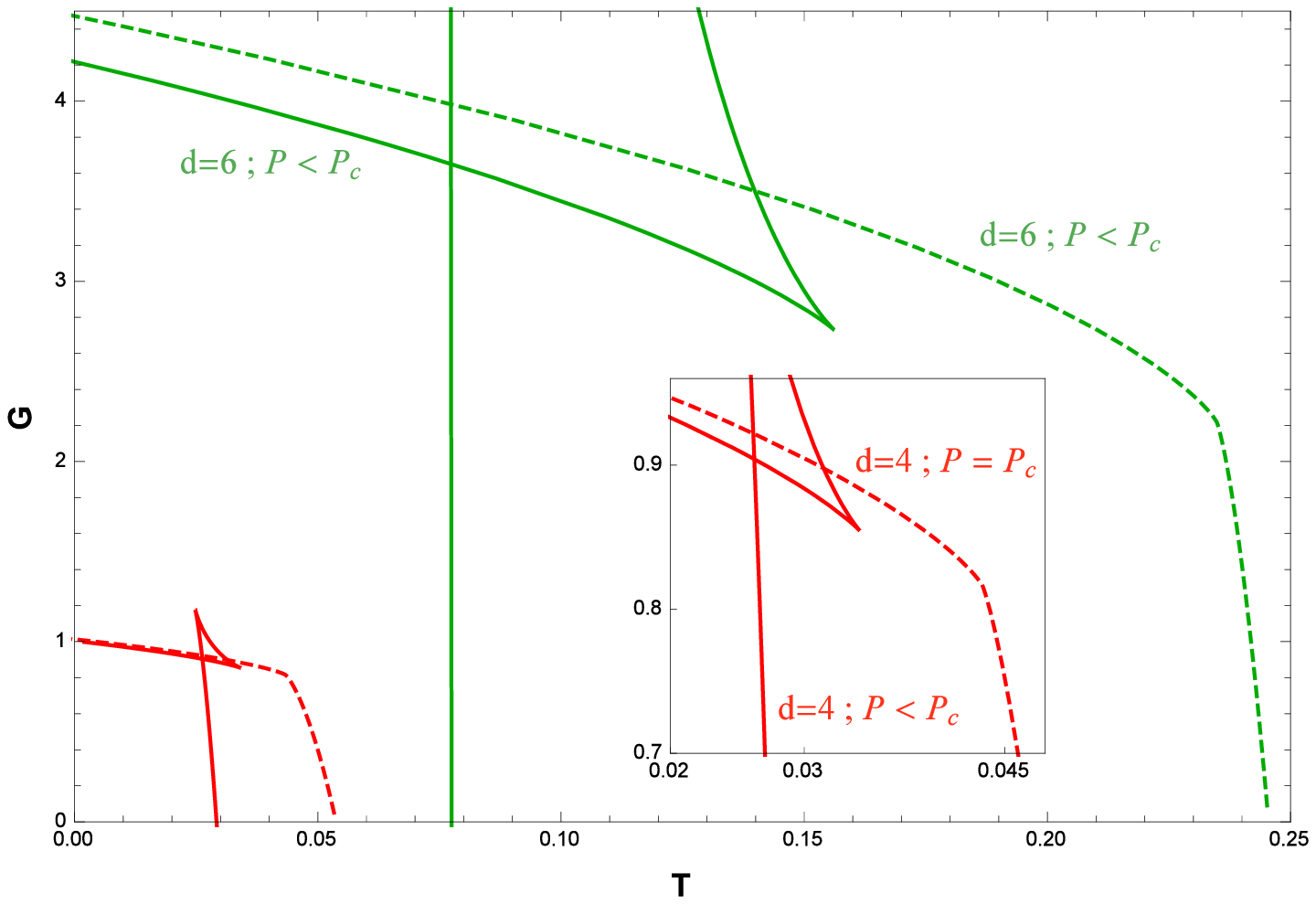}
			\end{tabbing}
			\vspace*{-.2cm} \caption{Left: the Hawking temperature as function of the black holes horizon $r_H$ in $d=4,6$. Right: the Gibbs free energy as function of the temperature $T$ in $d=4,6$. The two upper lines (purple) correspond to $d=6$ while the lower one (blue) correspond to $d=4$. The critical isobare $P=P_c$ is denoted by the dashed lines, the solid lines  shows the first-order transition $P<P_c$.}
			\label{fig2}
		\end{minipage}
\end{figure}

In Figure \ref{fig2}, we show the behavior of this transformation in $(T,r_H)$-diagram for $d=4$ and $d=6$.
From the left panel of Figure \ref{fig2}, we can see a small-large black hole phase transitions under  the critical regime $P<P_c$ (solid lines) and at $P=P_c$ (dashed lines), for which the coordinates of the critical point $(T_c,r_{H_{c}})$ are determined by the roots of the system:
 $\frac{\partial T}{\partial r_H}= \frac{\partial^2 T}{\partial r_H^2}=0$.
 The Gibbs free energy displays a characteristic swallow tail behavior, depicted in the right panel (solid lines), signaling a first-order SBH/LBH phase transition. The dashed line indicates the character of a second-order phase transition due a continuous variation of the free energy between SBH and LBH.

Table \ref{tab1} lists the frequencies of the quasinormal modes  of the massless scalar 
perturbation around small and large black holes for the first order phase transition where the pressure is fixed at the value $P=0.1P_c$. 

\begin{table}[!ht]
	\begin{center}
		\begin{minipage}{0.88\linewidth}
			{\center \tiny
				\begin{tabbing}
					\hspace{8.4cm}\=\kill
					
					\begin{tabular}{|p{0.8cm}|p{0.8cm}|p{0.9cm}p{1.3cm}|}	
						\multicolumn{4}{|c|}{$d=4$; $P=0.000331573$ and $T_{c}=0.0158676$}\\
						\hline
						\centering{$T$} & \centering{$r_{H}$} & \centering{$\omega_{r}$}  & \hspace{9pt}$\omega_{im}$\\ 
						\hline
						0.013 & 1.09711 & 0.140236 & -0.00361077 \\
						0.0135 & 1.10221 & 0.140229 & -0.00361519 \\
						0.014 & 1.10743 & 0.140222 & -0.00361991 \\
						0.0145 & 1.11279 & 0.140214 & -0.00362496 \\
						0.015 & 1.11829 & 0.140206 & -0.00363038 \\
						0.0155 & 1.12395 & 0.140197 & -0.0036362 \\
						\hline\hline
						0.016 & 17.1569 & 0.141771 & -0.127259 \\
						0.0165 & 18.3675 & 0.145464 & -0.136319 \\
						0.017 & 19.4967 & 0.149092 & -0.144749 \\
						0.0175 & 20.5692 & 0.152686 & -0.152739 \\
						0.018 & 21.5996 & 0.156262 & -0.160405 \\
						0.0185 & 22.5974 & 0.159829 & -0.167819 \\
						\hline 
					\end{tabular}
					\>\begin{tabular}{|p{0.8cm}|p{0.8cm}|p{0.9cm}p{1.3cm}|}	
						\multicolumn{4}{|c|}{$d=5$; $P=0.00205468$ and $T_{c}=0.0439405$}\\
						\hline
						\centering{$T$} & \centering{$r_{H}$} & \centering{$\omega_{r}$}  & \hspace{9pt}$\omega_{im}$\\ 
						\hline
						0.036 & 1.06452 & 0.35998  & -0.00328399 \\
						0.037 & 1.06699 & 0.359975 & -0.00328714 \\
						0.038 & 1.0695 & 0.35997 & -0.00329046 \\
						0.04 & 1.07464 & 0.35996 & -0.00329765 \\
						0.041 & 1.07729 & 0.359954 & -0.00330155 \\
						0.043 & 1.08273 & 0.359942 & -0.00331004 \\
						\hline\hline
						0.044 & 10.5599 & 0.420898 & -0.231517 \\
						0.045 & 11.2726 & 0.433701 & -0.249332 \\
						0.046 & 11.9157 & 0.445686 & -0.265327 \\
						0.047 & 12.5136 & 0.457162 & -0.280141 \\
						0.048 & 13.0794 & 0.46829 & -0.294112 \\
						0.049 & 13.621 & 0.479167 & -0.307447 \\
						\hline 
					\end{tabular} \\ \\
					\begin{tabular}{|p{0.8cm}|p{0.8cm}|p{0.9cm}p{1.3cm}|}	
						\multicolumn{4}{|c|}{$d=6$; $P=0.0058964$ and $T_{c}=0.0773752$}\\
						\hline
						\centering{$T$} & \centering{$r_{H}$} & \centering{$\omega_{r}$}  & \hspace{9pt}$\omega_{im}$\\ 
						\hline
						0.074 & 1.06156 & 0.60234 & -0.00223733 \\
						0.0745 & 1.06216 & 0.602339 & -0.00223825 \\
						0.075 & 1.06277 & 0.602338 & -0.00223919 \\
						0.0755 & 1.06338 & 0.602337 & -0.00224014 \\
						0.076 & 1.06399 & 0.602336 & -0.0022411 \\
						0.077 & 1.06523 & 0.602333 & -0.00224309 \\
						\hline\hline
						0.0775 & 8.21528 & 0.731459 & -0.28622 \\
						0.078 & 8.41977 & 0.739944 & -0.295132 \\
						0.0785 & 8.61178 & 0.748027 & -0.30348 \\
						0.079 & 8.79399 & 0.755798 & -0.311384 \\
						0.0795 & 8.96824 & 0.763316 & -0.318927 \\
						0.08 & 9.13587 & 0.770627 & -0.326169 \\
						\hline 
						
					\end{tabular}\>\begin{tabular}{|p{0.8cm}|p{0.8cm}|p{0.9cm}p{1.3cm}|}	
					\multicolumn{4}{|c|}{$d=7$; $P=0.0122899$ and $T_{c}=0.114151$}\\
					\hline
					\centering{$T$} & \centering{$r_{H}$} & \centering{$\omega_{r}$}  & \hspace{9pt}$\omega_{im}$\\ 
					\hline
					0.106 & 1.04847 & 0.85745 & -0.00138892 \\
					0.107 & 1.04913 & 0.857449 & -0.00138971 \\
					0.11 & 1.05116 & 0.857446 & -0.00139219 \\
					0.112 & 1.05254 & 0.857444 & -0.00139395 \\
					0.113 & 1.05324 & 0.857443 & -0.00139486 \\
					0.114 & 1.05394 & 0.857442 & -0.0013958 \\
					\hline\hline
					0.115 & 7.20019 & 1.05914 & -0.32404 \\
					0.116 & 7.4564 & 1.07735 & -0.339388 \\
					0.117 & 7.69002 & 1.09429 & -0.353327 \\
					0.118 & 7.90753 & 1.11033 & -0.366262 \\
					0.119 & 8.11287 & 1.12569 & -0.378435 \\
					0.12 & 8.30858 & 1.14053 & -0.390003 \\
					\hline 
				\end{tabular}\\ \\
				\begin{tabular}{|p{0.8cm}|p{0.8cm}|p{0.9cm}p{1.3cm}|}	
					\multicolumn{4}{|c|}{$d=8$; $P=0.021516$ and $T_{c}=0.153223$}\\
					\hline
					\centering{$T$} & \centering{$r_{H}$} & \centering{$\omega_{r}$}  & \hspace{9pt}$\omega_{im}$\\ 
					\hline
					0.147 & 1.04342 & 1.1213025 & -0.000836846 \\
					0.148 & 1.04386 & 1.1213020 & -0.000837338 \\
					0.15 & 1.04475 & 1.1213009 & -0.000838189 \\
					0.151 & 1.0452 & 1.1213004 & -0.000838626 \\
					0.152 & 1.04565 & 1.1212999 & -0.000839073 \\
					0.153 & 1.0461 & 1.1212994 & -0.000839529 \\
					\hline\hline
					0.154 & 6.40615 & 1.3714 & -0.333767 \\
					0.155 & 6.60836 & 1.39186 & -0.348736 \\
					0.156 & 6.79092 & 1.41068 & -0.362201 \\
					0.157 & 6.95969 & 1.42836 & -0.374608 \\
					0.158 & 7.11811 & 1.44519 & -0.38622 \\
					0.16 & 7.41211 & 1.47698 & -0.407682 \\
					\hline 
				\end{tabular} \> \begin{tabular}{|p{0.8cm}|p{0.8cm}|p{0.9cm}p{1.3cm}|}	
				\multicolumn{4}{|c|}{$d=9$; $P=0.0337696$ and $T_{c}=0.19396$}\\
				\hline
				\centering{$T$} & \centering{$r_{H}$} & \centering{$\omega_{r}$}  & \hspace{9pt}$\omega_{im}$\\ 
				\hline
				0.186 & 1.038 & 1.3917389 & -0.000495416 \\
				0.187 & 1.03831 & 1.3917386 & -0.00049562 \\
				0.189 & 1.03893 & 1.3917379 & -0.000496039 \\
				0.19 & 1.03924 & 1.3917376 & -0.000496254 \\
				0.192 & 1.03987 & 1.3917370 & -0.000496693 \\
				0.193 & 1.04018 & 1.3917366 & -0.000496918 \\
				\hline\hline
				0.194 & 5.75112 & 1.66464 & -0.32498 \\
				0.195 & 5.93915 & 1.68902 & -0.341193 \\
				0.196 & 6.10246 & 1.71065 & -0.355221 \\
				0.197 & 6.24999 & 1.73053 & -0.367852 \\
				0.199 & 6.51421 & 1.76691 & -0.390373 \\
				0.2 & 6.63541 & 1.78391 & -0.400662 \\
				\hline 
			\end{tabular}\\ \\
			\begin{tabular}{|p{0.8cm}|p{0.8cm}|p{0.9cm}p{1.3cm}|}	
				\multicolumn{4}{|c|}{$d=10$; $P=0.0491927$ and $T_{c}=0.235953$}\\
				\hline
				\centering{$T$} & \centering{$r_{H}$} & \centering{$\omega_{r}$}  & \hspace{9pt}$\omega_{im}$\\ 
				\hline
				0.227 & 1.03404 & 1.6676469 & -0.000292273 \\
				0.23 & 1.03473 & 1.6676445 & -0.000292587 \\
				0.232 & 1.03519 & 1.6676429 & -0.000292802 \\
				0.233 & 1.03543 & 1.6676421 & -0.000292912 \\
				0.234 & 1.03566 & 1.6676413 & -0.000293024 \\
				0.235 & 1.03589 & 1.6676405 & -0.000293137 \\
				\hline\hline
				0.236 & 5.40433 & 1.97324 & -0.324694 \\
				0.237 & 5.56862 & 1.99944 & -0.340532 \\
				0.238 & 5.70992 & 2.02246 & -0.354105 \\
				0.239 & 5.83679 & 2.04349 & -0.366254 \\
				0.24 & 5.95353 & 2.06313 & -0.3774 \\
				0.242 & 6.16581 & 2.09951 & -0.39759 \\
				\hline 
			\end{tabular} \> \begin{tabular}{|p{0.8cm}|p{0.8cm}|p{0.9cm}p{1.3cm}|}	
			\multicolumn{4}{|c|}{$d=11$; $P=0.0678928$ and $T_{c}=0.278925$}\\
			\hline
			\centering{$T$} & \centering{$r_{H}$} & \centering{$\omega_{r}$}  & \hspace{9pt}$\omega_{im}$\\  
			\hline
			0.271 & 1.03123 & 1.9497594 & -0.000176416 \\
			0.272 & 1.0314 & 1.9497558 & -0.000176466 \\
			0.274 & 1.03176 & 1.9497487 & -0.000176568 \\
			0.276 & 1.03212 & 1.9497416 & -0.000176673 \\
			0.277 & 1.0323 & 1.9497380 & -0.000176727 \\
			0.278 & 1.03248 & 1.9497344 & -0.000176781 \\
			\hline\hline
			0.279 & 5.14764 & 2.28145 & -0.322498 \\
			0.28 & 5.29436 & 2.30916 & -0.337921 \\
			0.281 & 5.41956 & 2.33331 & -0.351036 \\
			0.282 & 5.53144 & 2.35526 & -0.362723 \\
			0.283 & 5.63404 & 2.37569 & -0.37341 \\
			0.284 & 5.7297 & 2.39498 & -0.383349 \\
			\hline 
		\end{tabular}
	\end{tabbing}
	\caption{The quasinormal frequencies of the massless scalar perturbation as a function of  the black holes temperature. The upper part, above the horizontal line is for the small black hole phase while the lower part is for the large one.} }
\label{tab1}
\end{minipage}
\end{center}
\end{table}

One should keep in mind that, unlike the ordinary normal modes, the QNMs decay at certain rates. Hence, having complex frequencies, the real and imaginary parts can help to gain some insight on the oscillations and decays of the black hole.
In the case of  the small black hole phase, for any dimension, we note from table $1$ that the black hole
becomes smaller and smaller  when the temperature decreases from the 
phase transition critical point $T_c$. In this process, the real part of the QNMs frequencies varies slightly, 
while the absolute value of the imaginary part decreases which indicates that the absorption performance of the black hole is reduced whatever the spacetime dimension $d$. This result is consistent with the overall analysis reported in \cite{base,zhuWang}.

For large black hole phase, for any dimension $d$,  we see that by increasing the temperature
from the critical value $T_c$, the black hole becomes larger.
In this case, the real part as well as the absolute value of the imaginary part
of quasinormal frequencies increase. 
Furthermore, the massless scalar perturbation outside the black hole performs more oscillations but decays faster. These observations are in agreement with the results found in
  \cite{base,WangMolina}.

\begin{figure}[!ht]
	\hspace*{-0.8cm}\begin{minipage}{1.13\linewidth}	
		\begin{tabbing}
			\hspace{4.5cm}\=\hspace{4.5cm}\=\hspace{4.5cm}\=\kill
			\includegraphics[width=4.5cm,height=3.cm]{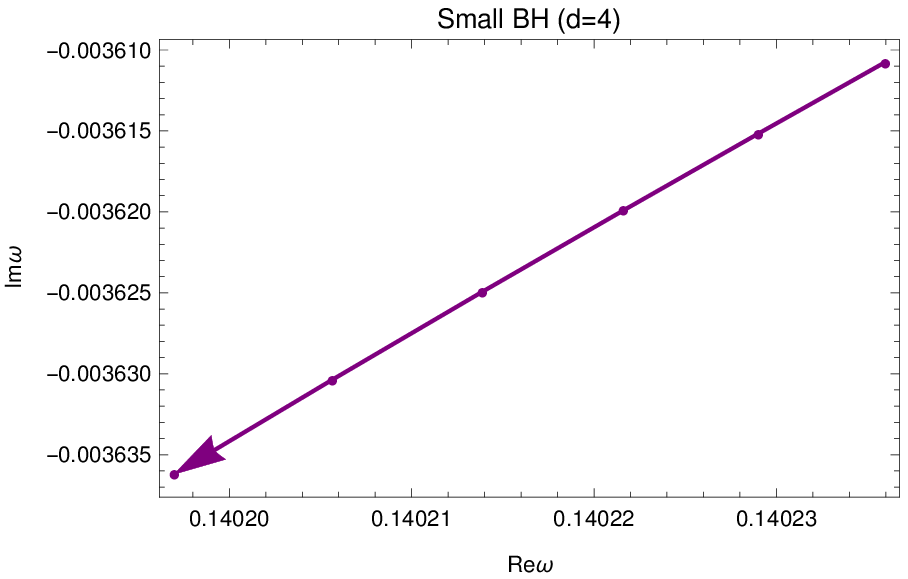}\> \includegraphics[width=4.5cm,height=3.cm]{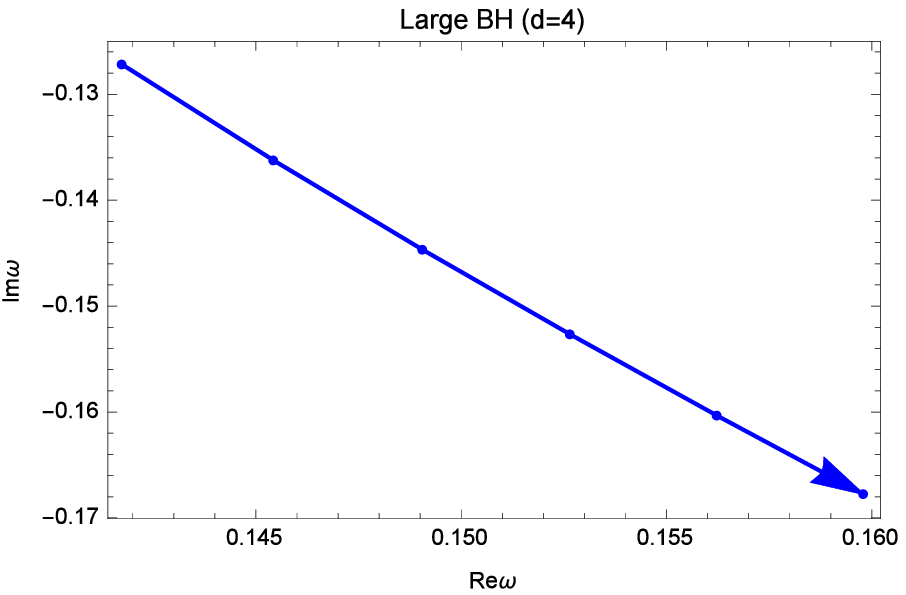} \>
			\includegraphics[width=4.5cm,height=3.cm]{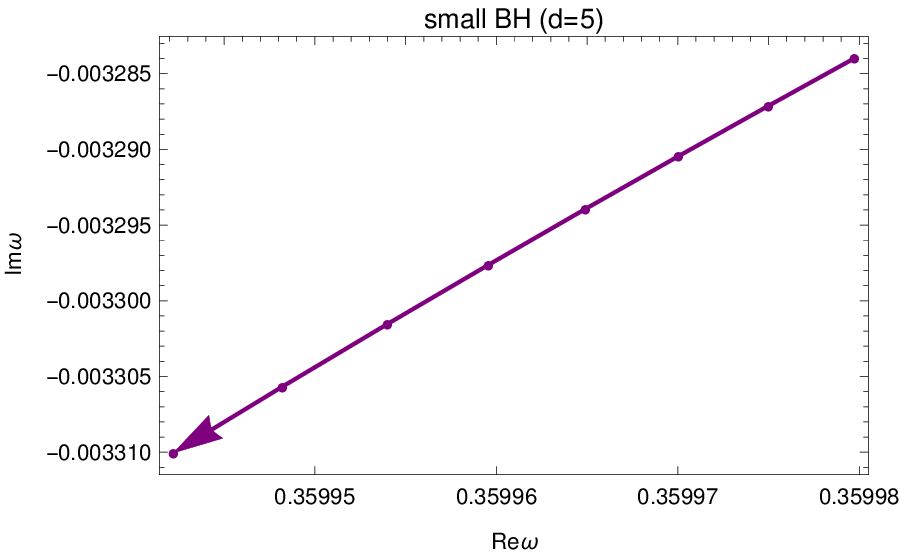}\> \includegraphics[width=4.5cm,height=3.cm]{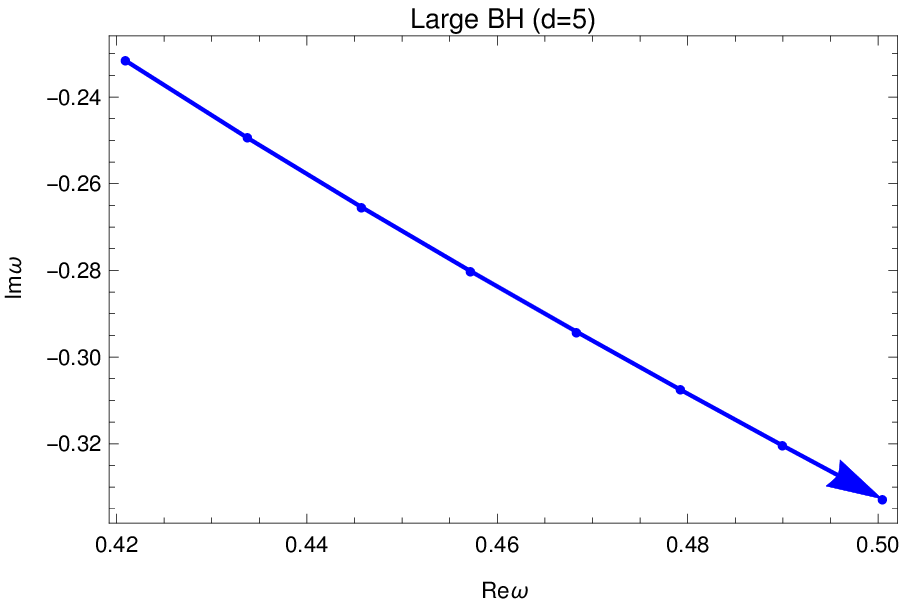} \\
			\includegraphics[width=4.5cm,height=3.cm]{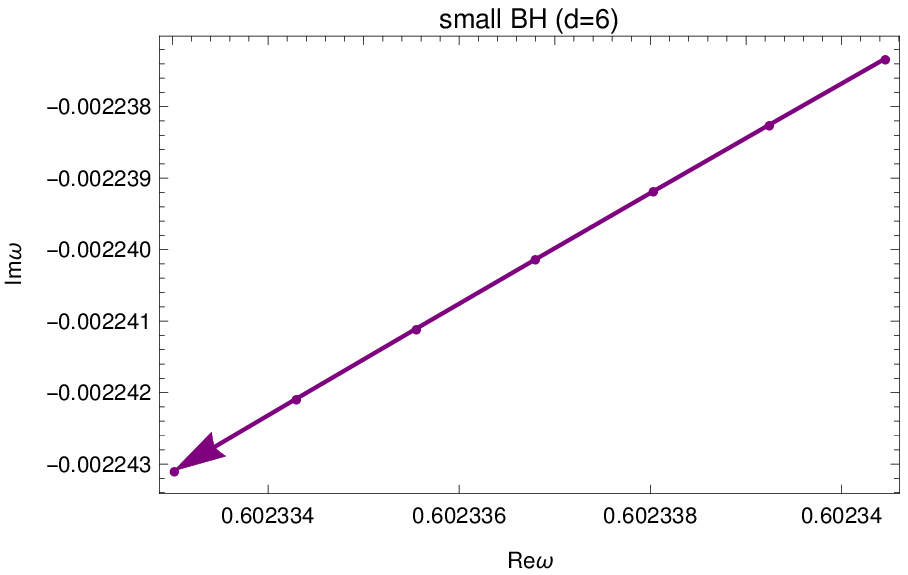}\> \includegraphics[width=4.5cm,height=3.cm]{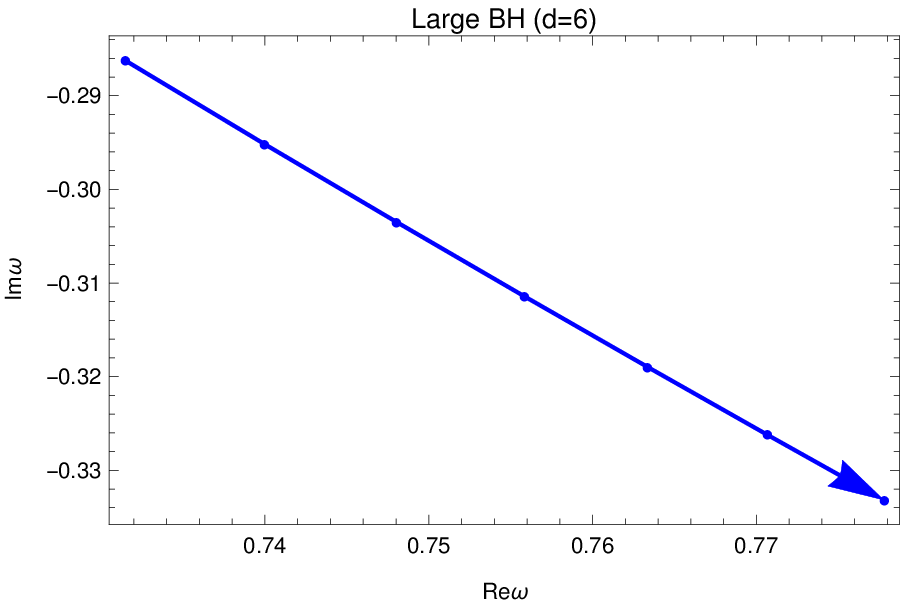} \>
			\includegraphics[width=4.5cm,height=3.cm]{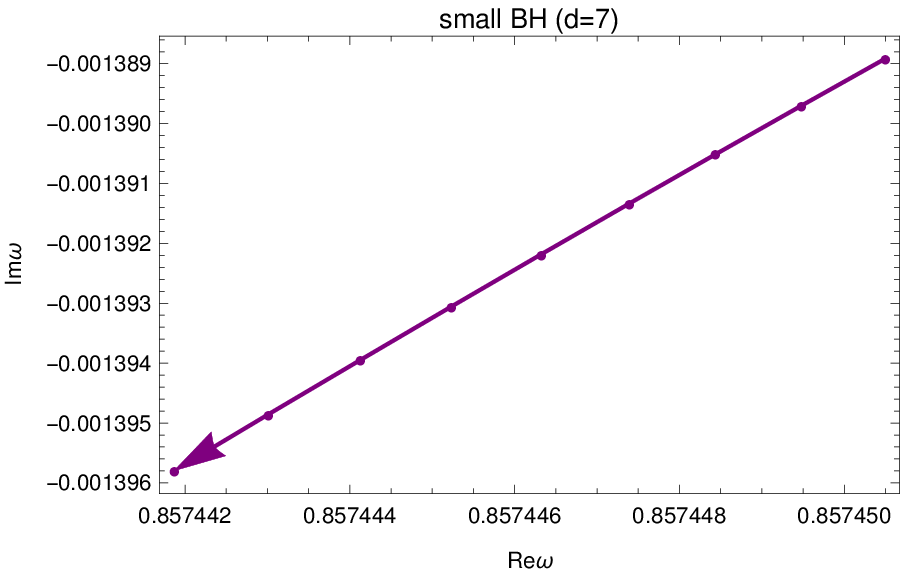}\> \includegraphics[width=4.5cm,height=3.cm]{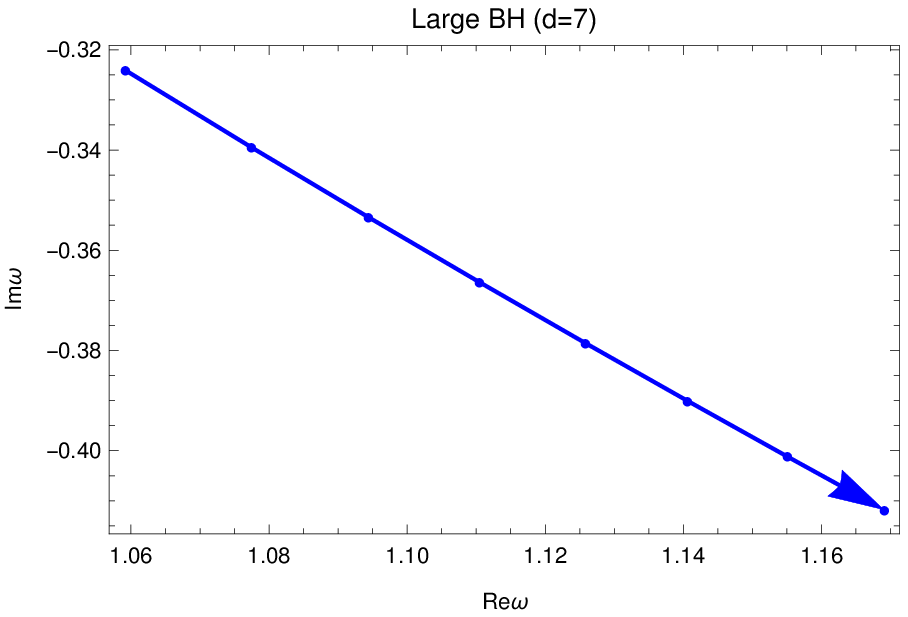} \\
			\includegraphics[width=4.5cm,height=3.cm]{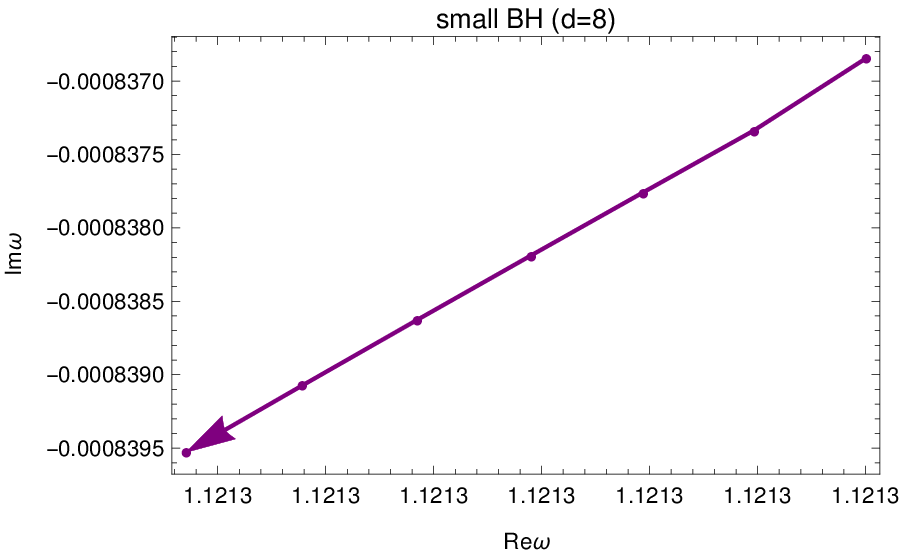}\> \includegraphics[width=4.5cm,height=3.cm]{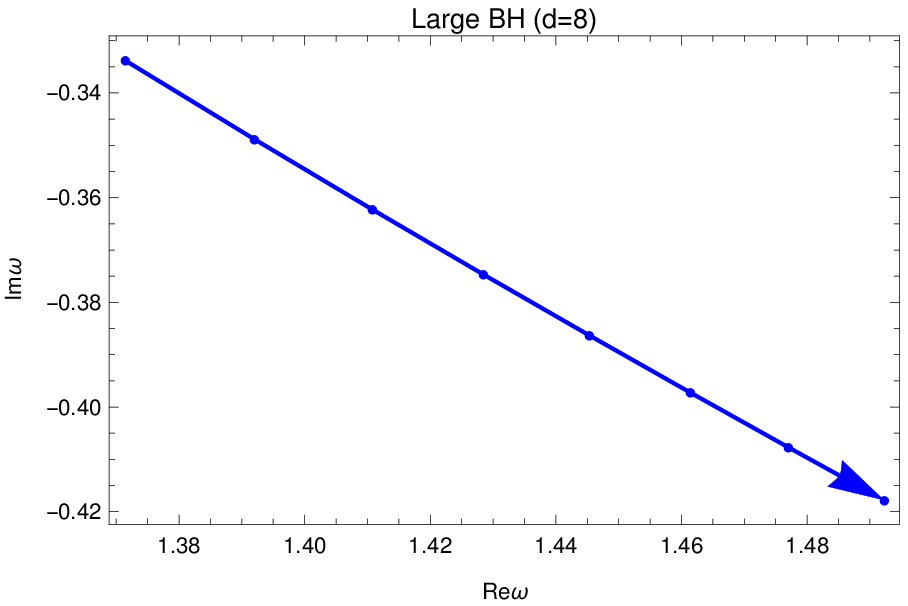} \>
			\includegraphics[width=4.5cm,height=3.cm]{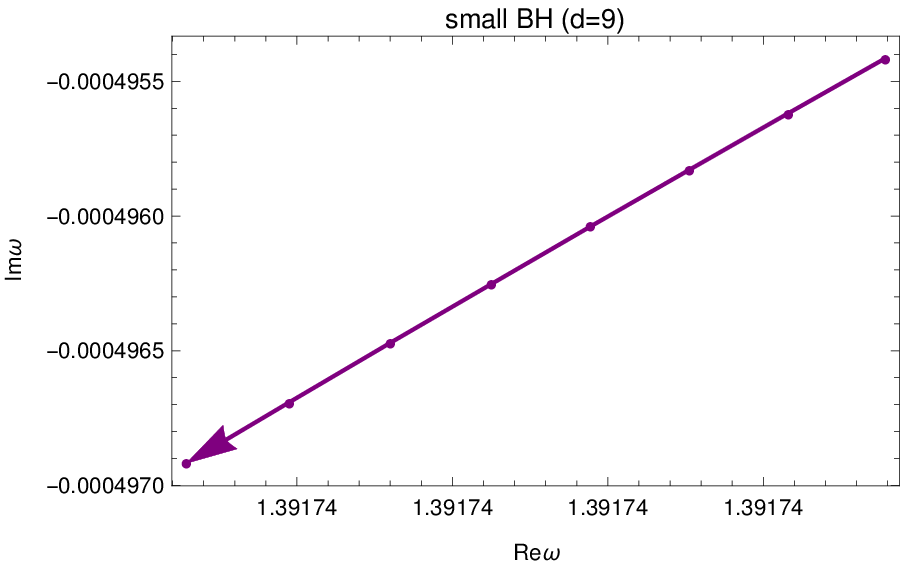}\> \includegraphics[width=4.5cm,height=3.cm]{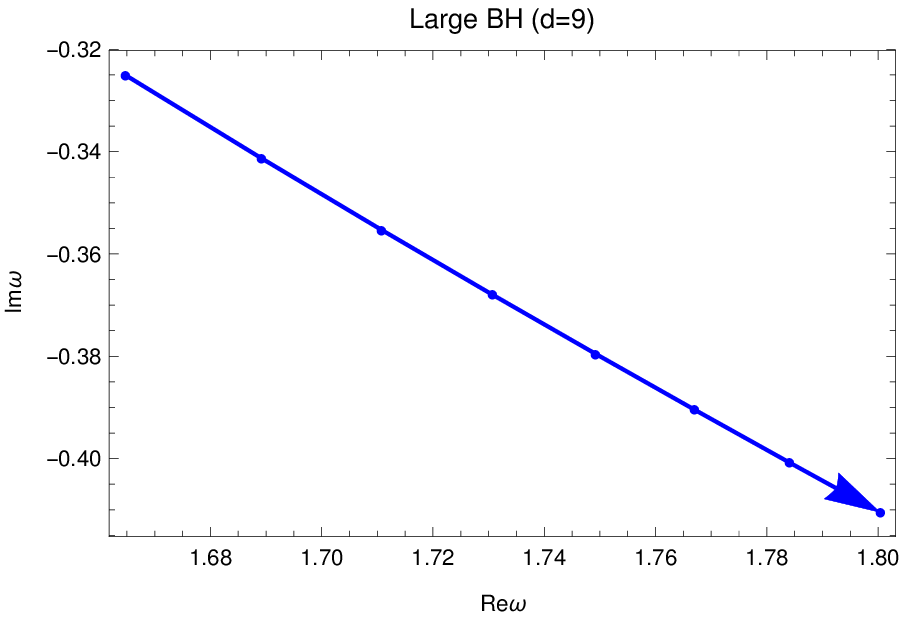} \\
			\includegraphics[width=4.5cm,height=3.cm]{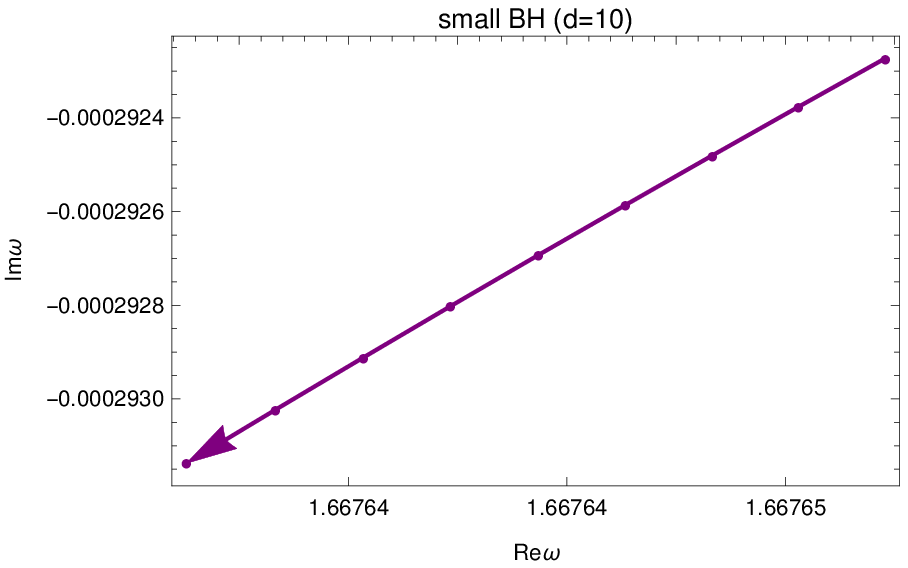}\> \includegraphics[width=4.5cm,height=3.cm]{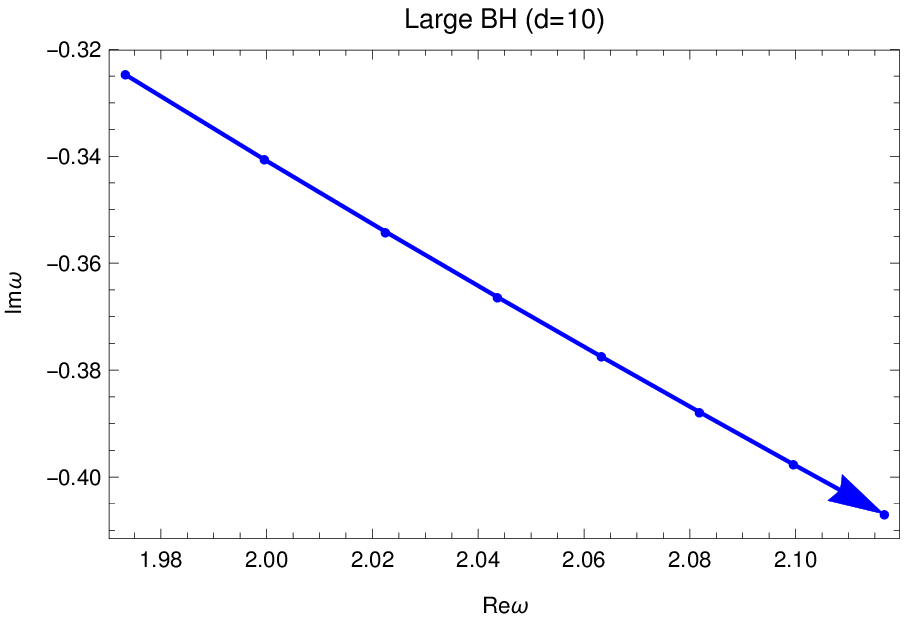} \>
			\includegraphics[width=4.5cm,height=3.cm]{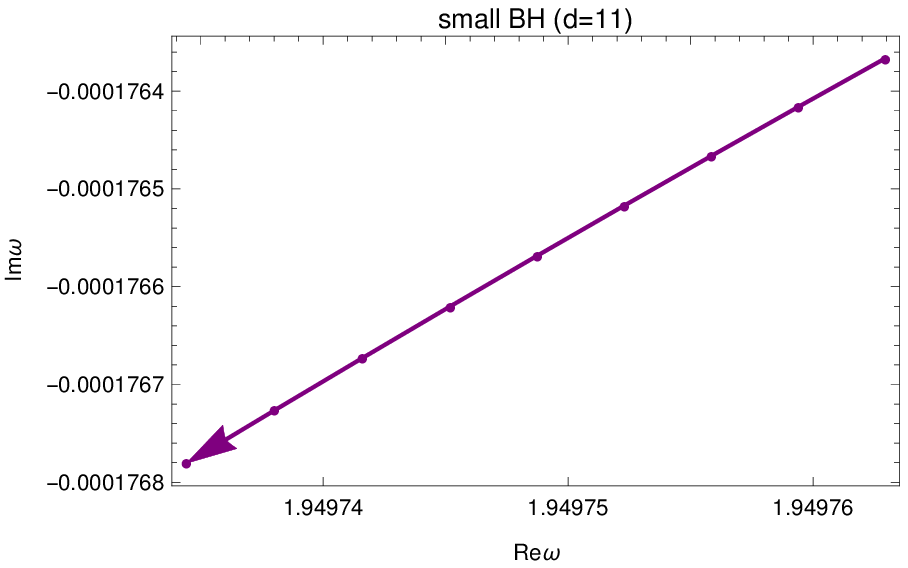}\> \includegraphics[width=4.5cm,height=3.cm]{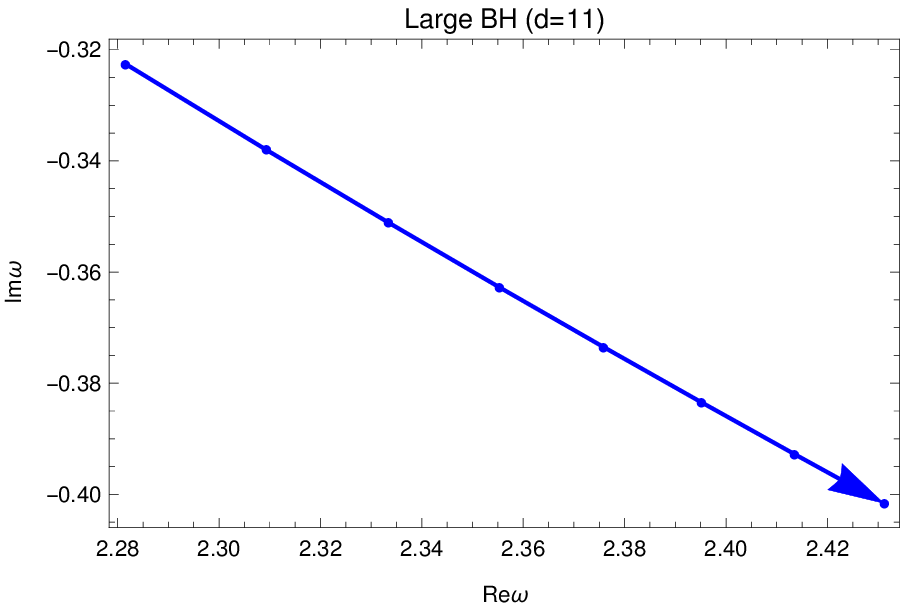} \\
		\end{tabbing}
		\vspace*{0cm} \caption{The behavior of the quasinormal modes for small and large black holes in the complex-$\omega$ plane for  $4\leq d\leq 11$. Increase of the black hole size is shown by the arrows for the isobaric process.}
		\label{fig3}
	\end{minipage}
\end{figure}

Figure \ref{fig3} illustrates the  quasinormal frequencies for small and
large black hole phases. Increase in the black hole' size is indicated by the arrows. 

From Figure \ref{fig3}, we see different slopes of the quasinormal frequencies in the massless scalar perturbations with different phases of the small and large black holes. The SBH and  LBH transform to each other through a phase transition. All these features revealed by dynamical perturbations are concording with  the general picture of thermodynamic phase transitions between small-large RN-AdS black holes.

\begin{figure}[!ht]
	\hspace*{-0.8cm}\begin{minipage}{1.13\linewidth}
		\begin{tabbing}
			\hspace{4.5cm}\=\hspace{4.5cm}\=\hspace{4.5cm}\=\kill
			\includegraphics[width=4.5cm,height=3.cm]{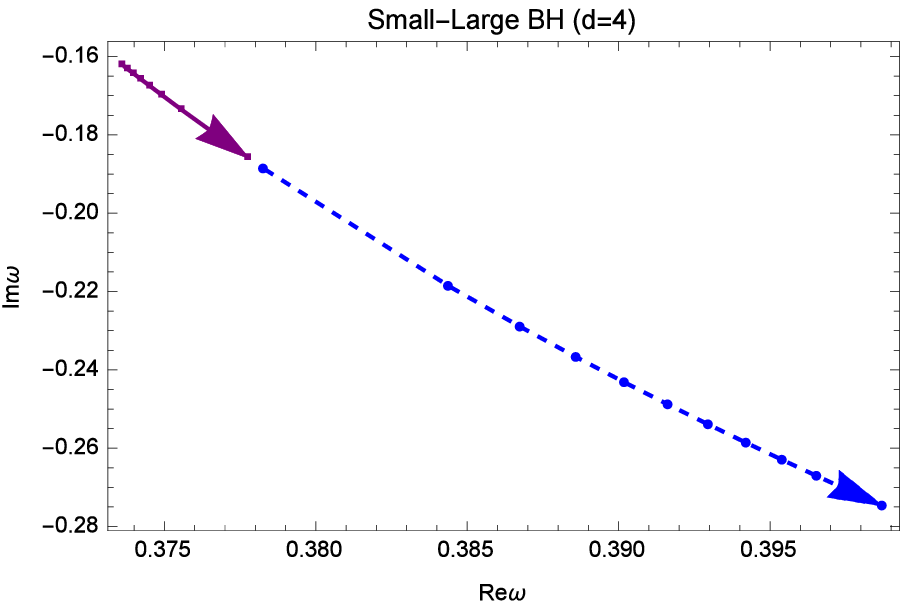}\> \includegraphics[width=4.5cm,height=3.cm]{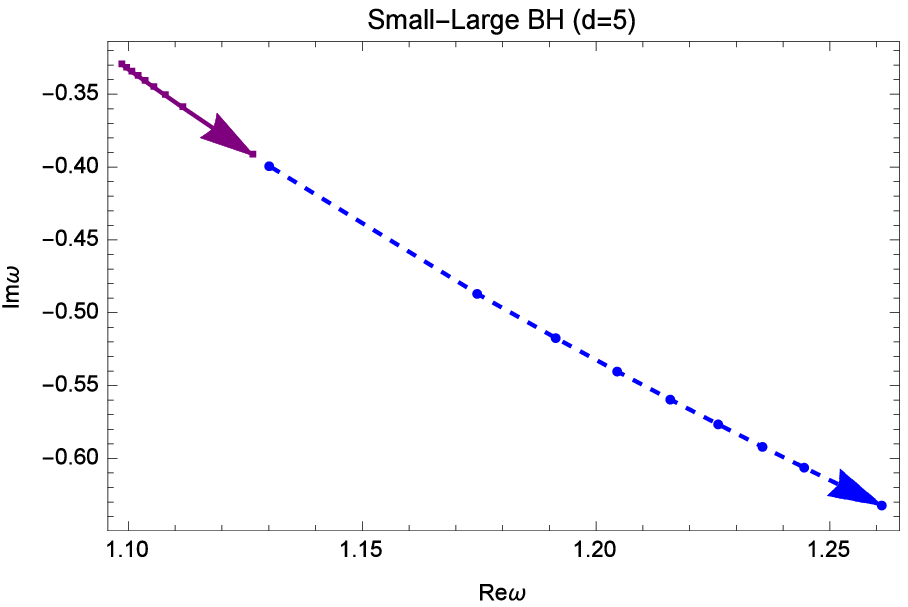} \>
			\includegraphics[width=4.5cm,height=3.cm]{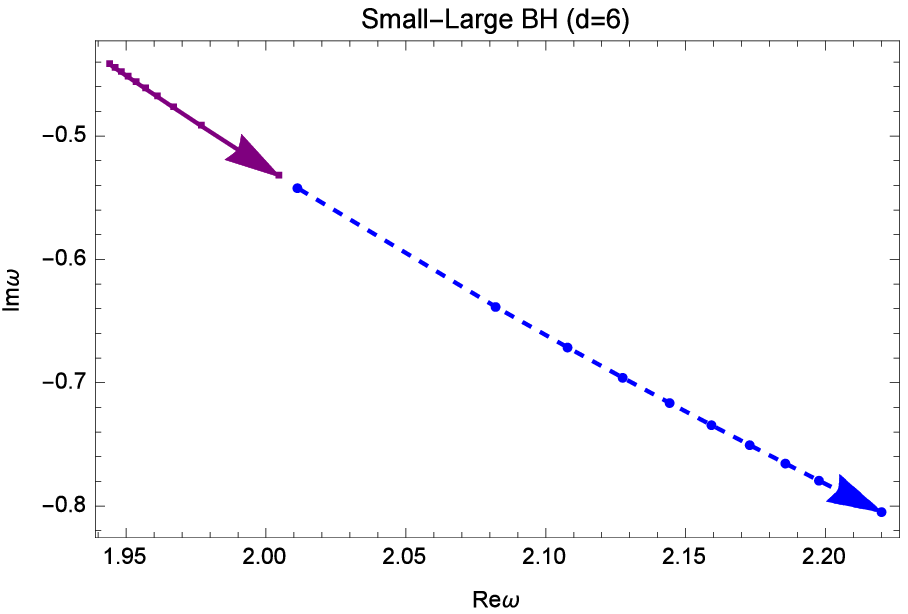}\> \includegraphics[width=4.5cm,height=3.cm]{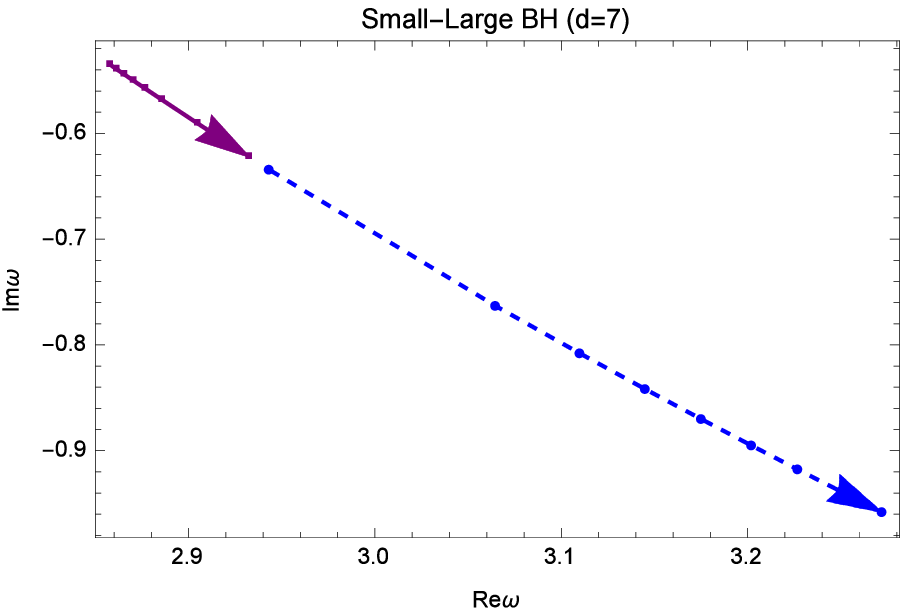} \\
			\includegraphics[width=4.5cm,height=3.cm]{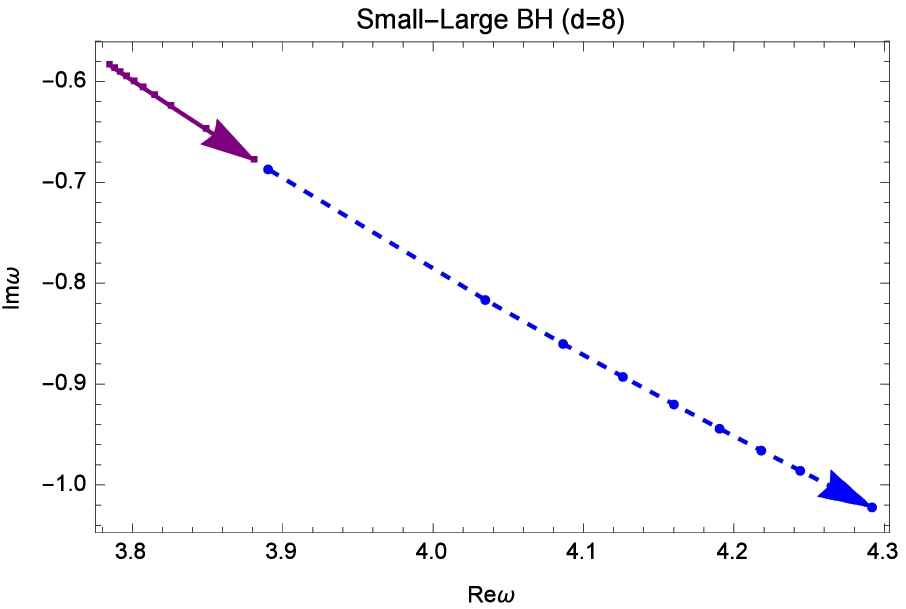}\> \includegraphics[width=4.5cm,height=3.cm]{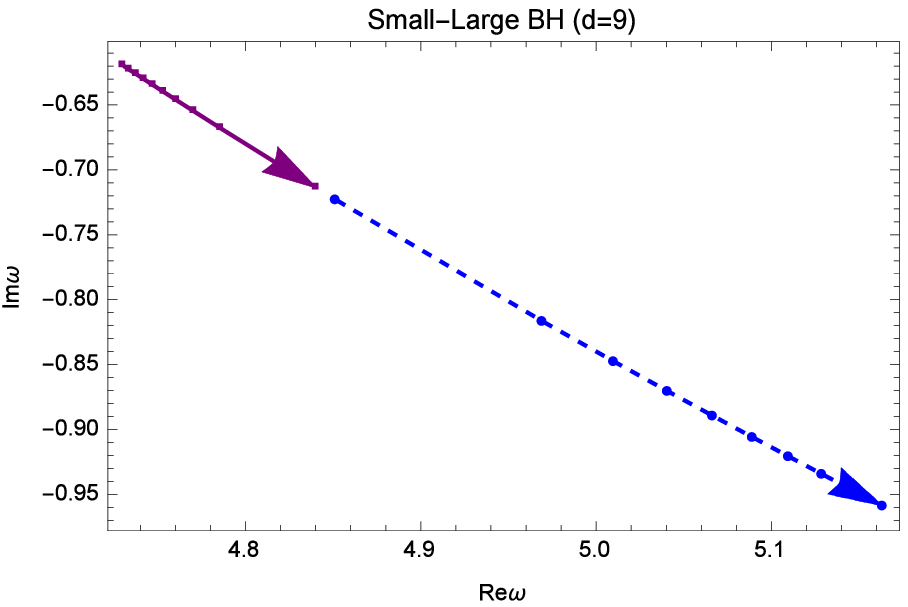} \>
			\includegraphics[width=4.5cm,height=3.cm]{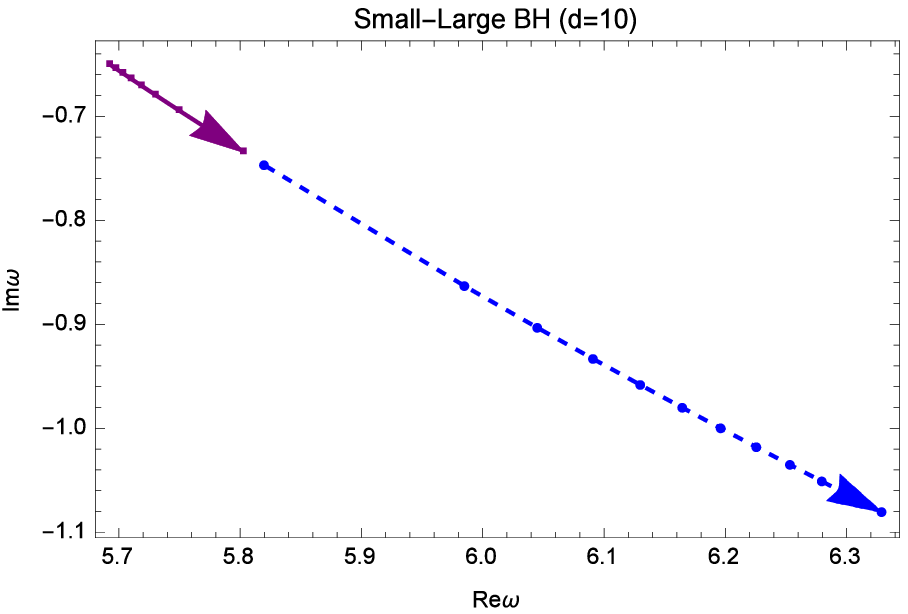}\> \includegraphics[width=4.5cm,height=3.cm]{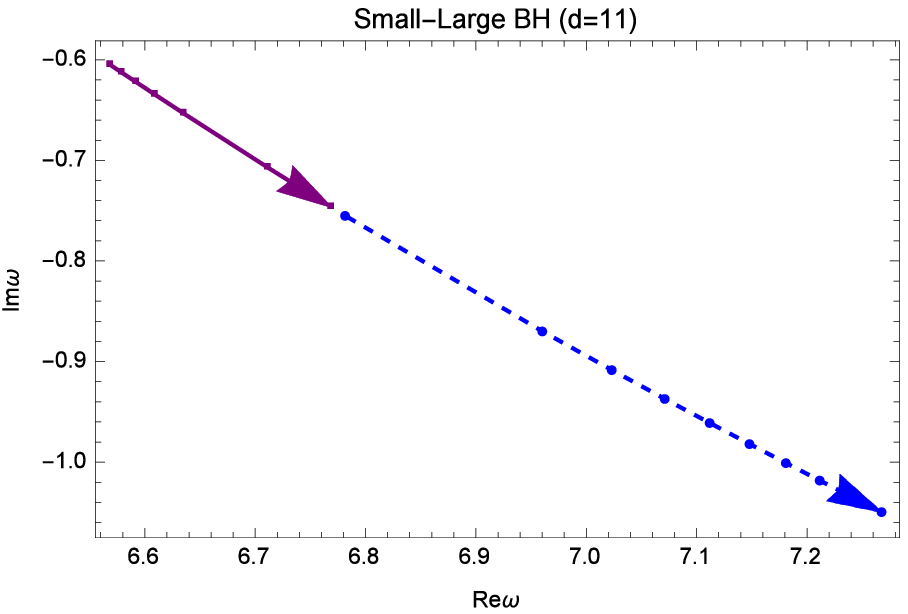} \\
		\end{tabbing}
		\vspace*{0cm} \caption{The behavior of the quasinormal modes for small (solid) and large (dashed) black holes in the isobaric second order  phase transition for a $4\leq d\leq 11$. Increase of the black hole size is shown by the arrows. }\label{fig7}
	\end{minipage}
\end{figure}

At the isobaric phase transition critical point $P=P_c$, the situation is different. The quasinormal frequencies for small and large black hole phases are plotted in Figure \ref{fig7}  for $4\leq d \leq 11$. We can see that when phase transitions are realized, both the SBH and LBH QNMs have the same behavior when the horizon increases, generalizing to higher spacetime dimensions the propriety found out in  $d=4$ \cite{base}.
\section{QNM behaviors in the isothermal phase transition\label{isothermal}}
\begin{figure}[!ht]
	\hspace*{0.8cm}	\begin{minipage}{0.88\linewidth}
			\begin{tabbing}
				\hspace{7.5cm}\=\kill
				\includegraphics[scale=.46]{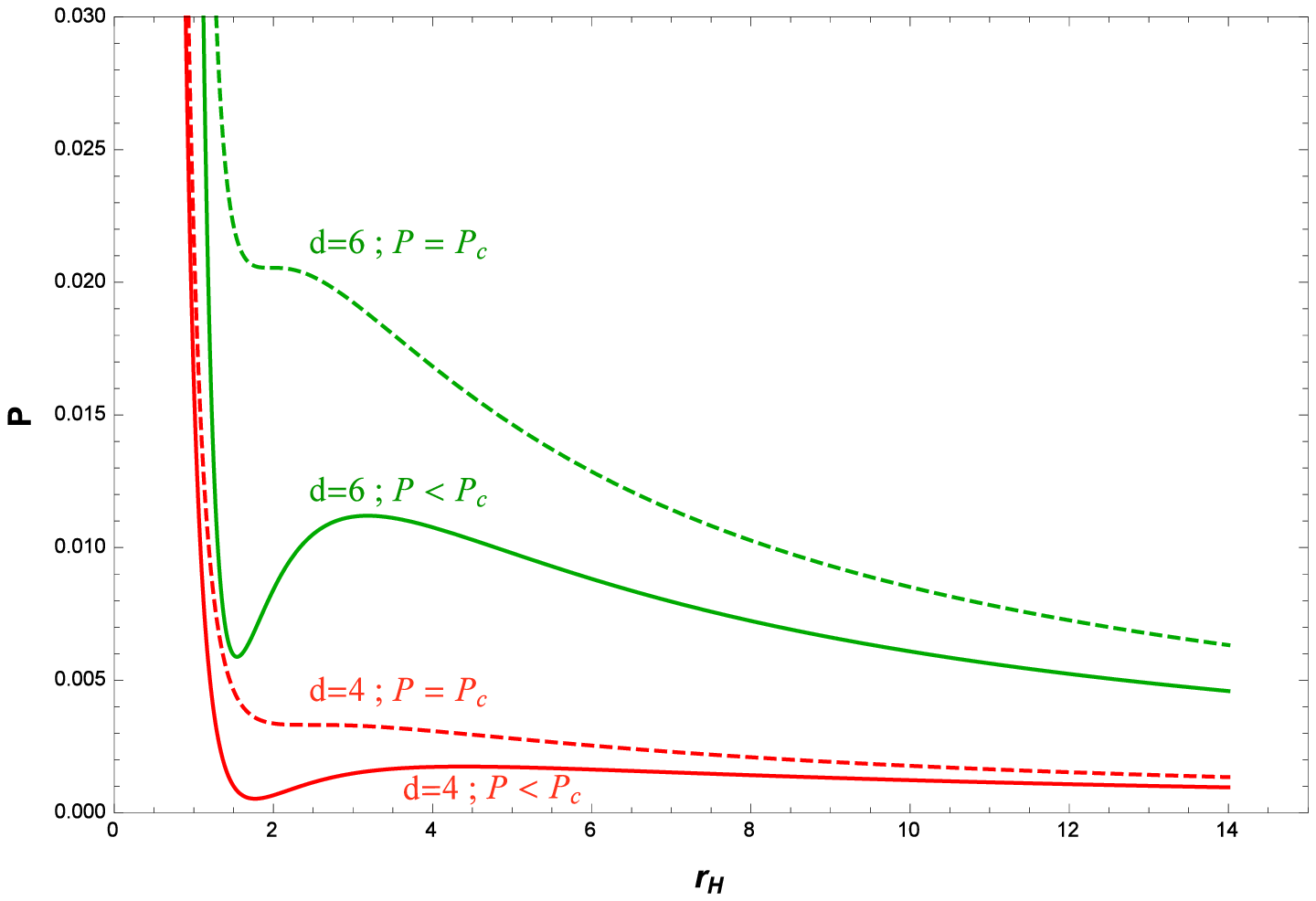}\> \includegraphics[scale=.46]{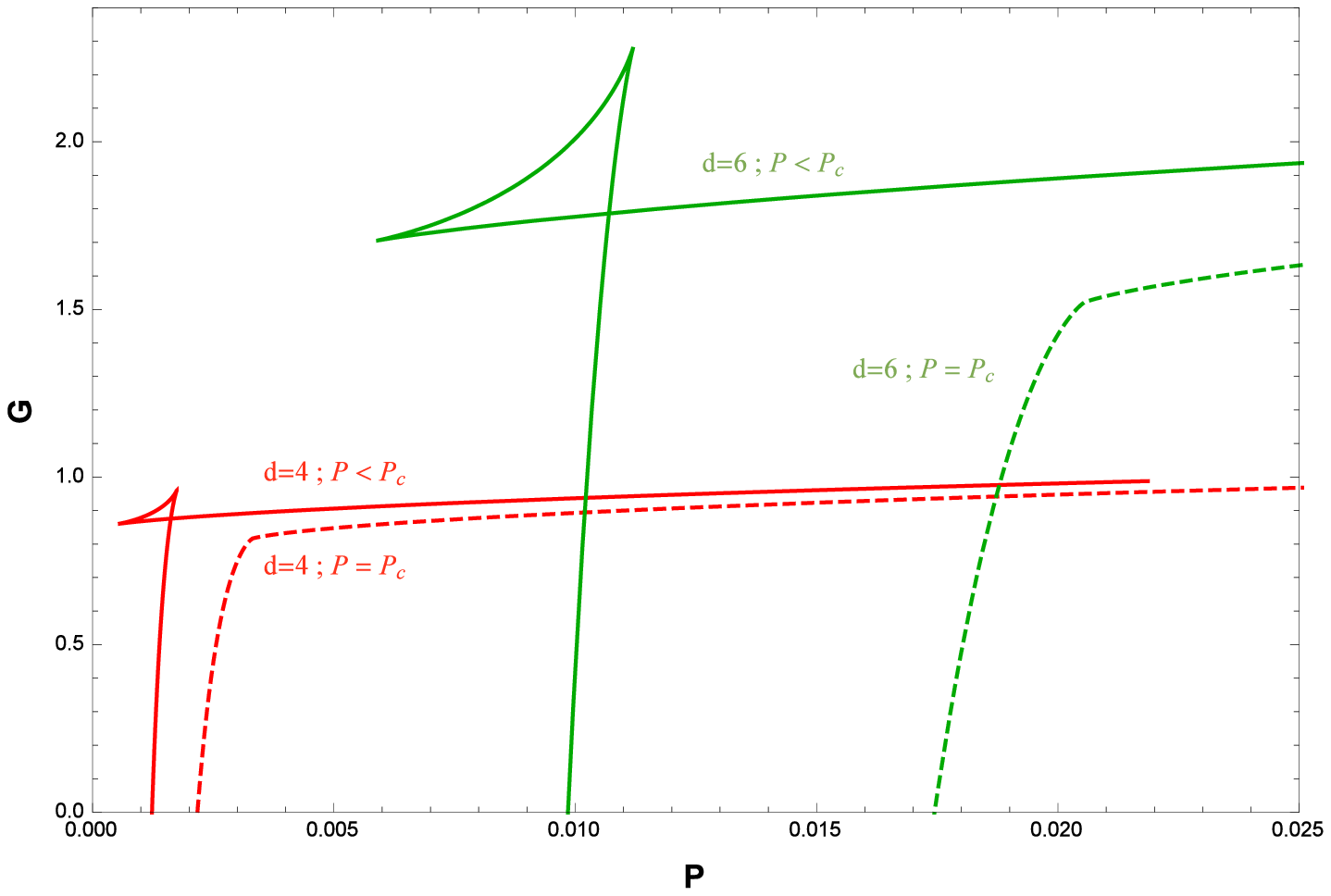} 
			\end{tabbing}
			\vspace*{-.2cm} \caption{Left panel: pressure $P$ as function of the horizon radius $r_H$ for isothermal process. Right panel: Gibbs free energy as function of the pressure for $d=4,6$. The two upper lines (purple) correspond to $d=6$ while the lower two  lines (blue) correspond to $d=4$. The critical isotherm $T=T_c$ is denoted by the dashed lines, the solid lines  shows the first-order transition $T<T_c$.}
			\label{fig4}
		\end{minipage}
\end{figure}

In this section, we fix the temperature $T$  and made the study in the $(P,r_H)$-diagram. In Figure \ref{fig4},
we plot the pressure of the black hole in terms of the radius of event horizon and the Gibbs free energy. 

The associated $(P,r_H)$-diagram
for $d=4$ black hole is displayed in Figure \ref{fig4} (left panel). Obviously, for $T < T_c$ there is a small-large black hole phase transition in the system. This qualitative behavior persists in higher dimensions, as illustrated for $d= 6$. In the right panel the behavior of $G$ for dimensions $d=4$ and $d=6$ is depicted in Figure \ref{fig4}. Similarly to Figure \ref{fig2}, the characteristic $1st$-order phase transition behavior shows up.
 
 Table \ref{tab2} displays the frequencies of the quasinormal modes for small and large black hole phase in the isothermal precess for different dimensions, with $T=0.74T_c$. The data above (below) the horizontal line are for the small (large) black hole phase, respectively.
\begin{table}[!ht]\tiny
	\begin{center}
		\begin{minipage}{0.88\linewidth}
			\begin{tabbing}
				\hspace{8.4cm}\=\kill
				
				\begin{tabular}{|p{0.8cm}|p{0.8cm}|p{0.9cm}p{1.3cm}|}	
					\multicolumn{4}{|c|}{$d=4$; $P_{c}=0.00157088$ and $T=0.0320542$}\\
					\hline
					\centering{$P$} & \centering{$r_{H}$} & \centering{$\omega_{r}$}  & \hspace{9pt}$\omega_{im}$\\ 
					\hline
					0.0023 & 1.32801 & 0.313337 & -0.0766033 \\
					0.0021 & 1.34336 & 0.301581 & -0.0688084 \\
					0.002 & 1.35161 & 0.295472 & -0.0649205 \\
					0.0018 & 1.36945 & 0.282723 & -0.0571742 \\
					0.0017 & 1.37917 & 0.276049 & -0.0533207 \\
					0.0016 & 1.38951 & 0.269149 & -0.0494843 \\
					\hline\hline
					0.00156 & 6.36627 & 0.286755 & -0.222296 \\
					0.0014 & 7.32754 & 0.285987 & -0.239667 \\
					0.0013 & 8.31838 & 0.284555 & -0.253537 \\
					0.0012 & 9.39777 & 0.282746 & -0.26541 \\
					0.0011 & 10.6133 & 0.280682 & -0.275946 \\
					0.0010 & 12.0182 & 0.278429 & -0.285503 \\
					\hline
				\end{tabular}
				\>\begin{tabular}{|p{0.8cm}|p{0.8cm}|p{0.9cm}p{1.3cm}|}	
					\multicolumn{4}{|c|}{$d=5$; $P_{c}=0.0103914$ and $T=0.0957523$}\\
					\hline
					\centering{$P$} & \centering{$r_{H}$} & \centering{$\omega_{r}$}  & \hspace{9pt}$\omega_{im}$\\ 
					\hline
					0.0135 & 1.23535 & 0.8608 & -0.138736 \\
					0.013 & 1.24149 & 0.845309 & -0.131794 \\
					0.012 & 1.2548 & 0.813519 & -0.118079 \\
					0.0115 & 1.26207 & 0.79718 & -0.111318 \\
					0.011 & 1.26981 & 0.780515 & -0.104628 \\
					0.0105 & 1.2781 & 0.763497 & -0.0980188 \\
					\hline\hline
					0.0045 & 14.074 & 0.944997 & -0.714101 \\
					0.004 & 16.1001 & 0.944984 & -0.728163 \\
					0.003 & 22.141 & 0.944314 & -0.754857 \\
					0.002 & 34.1599 & 0.942915 & -0.779934 \\
					0.001 & 70.1117 & 0.940907 & -0.803661 \\
					0.0005 & 141.947 & 0.939703 & -0.815083 \\
					\hline 
				\end{tabular} \\ \\
				\begin{tabular}{|p{0.8cm}|p{0.8cm}|p{0.9cm}p{1.3cm}|}	
					\multicolumn{4}{|c|}{$d=6$; $P_{c}=0.0306566$ and $T=0.173793$}\\
					\hline
					\centering{$P$} & \centering{$r_{H}$} & \centering{$\omega_{r}$}  & \hspace{9pt}$\omega_{im}$\\ 
					\hline
					0.037 & 1.18927 & 1.46326 & -0.166943 \\
					0.035 & 1.19798 & 1.42273 & -0.153541 \\
					0.034 & 1.20266 & 1.40213 & -0.14694 \\
					0.033 & 1.20759 & 1.38128 & -0.140413 \\
					0.032 & 1.21278 & 1.36018 & -0.133963 \\
					0.031 & 1.2183 & 1.3388 & -0.127597 \\
					\hline\hline
					0.03 & 3.55703 & 1.63219 & -0.626789 \\
					0.025 & 5.06739 & 1.70814 & -0.786255 \\
					0.02 & 6.97943 & 1.74585 & -0.891377 \\
					0.015 & 9.99367 & 1.77001 & -0.976436 \\
					0.01 & 15.8756 & 1.78665 & -1.05007 \\
					0.005 & 33.326 & 1.79836 & -1.11605 \\
					\hline
				\end{tabular}\>\begin{tabular}{|p{0.8cm}|p{0.8cm}|p{0.9cm}p{1.3cm}|}	
				\multicolumn{4}{|c|}{$d=7$; $P_{c}=0.0648614$ and $T=0.2602$}\\
				\hline
				\centering{$P$} & \centering{$r_{H}$} & \centering{$\omega_{r}$}  & \hspace{9pt}$\omega_{im}$\\ 
				\hline
				0.068 & 1.17485 & 1.98887 & -0.152482 \\
				0.067 & 1.17726 & 1.97383 & -0.148805 \\
				0.0665 & 1.17849 & 1.96627 & -0.146978 \\
				0.066 & 1.17975 & 1.9587 & -0.145159 \\
				0.0655 & 1.18103 & 1.9511 & -0.143348 \\
				0.065 & 1.18233 & 1.94348 & -0.141544 \\
				\hline\hline
				0.064 & 3.03158 & 2.37227 & -0.700765 \\
				0.059 & 3.6804 & 2.45839 & -0.822831 \\
				0.054 & 4.31594 & 2.51127 & -0.90842 \\
				0.049 & 5.0203 & 2.55064 & -0.978804 \\
				0.044 & 5.84492 & 2.5822 & -1.04021 \\
				0.039 & 6.85048 & 2.60853 & -1.0955 \\
				\hline 
			\end{tabular}\\ \\
			\begin{tabular}{|p{0.8cm}|p{0.8cm}|p{0.9cm}p{1.3cm}|}
				\multicolumn{4}{|c|}{$d=8$; $P_{c}=0.114605$ and $T=0.352107$}\\
				\hline
				\centering{$P$} & \centering{$r_{H}$} & \centering{$\omega_{r}$}  & \hspace{9pt}$\omega_{im}$\\   
				\hline
				0.117 & 1.15334 & 2.60132 & -0.154053 \\
				0.1162 & 1.15443 & 2.59209 & -0.152204 \\
				0.116 & 1.1547 & 2.58977 & -0.151743 \\
				0.1154 & 1.15553 & 2.58283 & -0.150363 \\
				0.115 & 1.1561 & 2.5782 & -0.149446 \\
				0.1148 & 1.15638 & 2.57588 & -0.148988 \\
				\hline\hline
				0.114 & 2.67852 & 3.10198 & -0.727098 \\
				0.109 & 3.05069 & 3.19286 & -0.827744 \\
				0.104 & 3.38124 & 3.25244 & -0.899785 \\
				0.099 & 3.71 & 3.29864 & -0.959385 \\
				0.089 & 4.41574 & 3.36992 & -1.05857 \\
				0.079 & 5.24526 & 3.42501 & -1.1421 \\
				\hline 
			\end{tabular} \> \begin{tabular}{|p{0.8cm}|p{0.8cm}|p{0.9cm}p{1.3cm}|}	
			\multicolumn{4}{|c|}{$d=9$; $P_{c}=0.180982$ and $T=0.447907$}\\
			\hline
			\centering{$P$} & \centering{$r_{H}$} & \centering{$\omega_{r}$}  & \hspace{9pt}$\omega_{im}$\\ 
			\hline
			0.183 & 1.13575 & 3.23913 & -0.154243 \\
			0.1824 & 1.13625 & 3.23358 & -0.1533 \\
			0.182 & 1.13659 & 3.22989 & -0.152672 \\
			0.1814 & 1.1371 & 3.22433 & -0.151733 \\
			0.1812 & 1.13727 & 3.22248 & -0.151421 \\
			0.181 & 1.13744 & 3.22063 & -0.151108 \\
			\hline\hline
			0.18 & 2.49165 & 3.84363 & -0.750582 \\
			0.175 & 2.73034 & 3.92879 & -0.83067 \\
			0.17 & 2.93771 & 3.98859 & -0.890562 \\
			0.16 & 3.33138 & 4.07696 & -0.98509 \\
			0.15 & 3.73361 & 4.14414 & -1.06217 \\
			0.14 & 4.16632 & 4.19932 & -1.12918 \\
			\hline 
		\end{tabular}\\ \\
		\begin{tabular}{|p{0.8cm}|p{0.8cm}|p{0.9cm}p{1.3cm}|}	
			\multicolumn{4}{|c|}{$d=10$; $P_{c}=0.264781$ and $T=0.546605$}\\
			\hline
			\centering{$P$} & \centering{$r_{H}$} & \centering{$\omega_{r}$}  & \hspace{9pt}$\omega_{im}$\\ 
			\hline
			0.267 & 1.12144 & 3.89469 & -0.153065 \\
			0.266 & 1.12199 & 3.88705 & -0.151936 \\
			0.2658 & 1.1221 & 3.88552 & -0.15171 \\
			0.2654 & 1.12232 & 3.88246 & -0.15126 \\
			0.265 & 1.12254 & 3.8794 & -0.15081 \\
			0.2648 & 1.12266 & 3.87787 & -0.150586 \\
			\hline\hline
			0.26 & 2.46955 & 4.63737 & -0.806927 \\
			0.25 & 2.75572 & 4.75123 & -0.90625 \\
			0.24 & 3.01582 & 4.8317 & -0.981443 \\
			0.23 & 3.27327 & 4.89615 & -1.04485 \\
			0.22 & 3.53769 & 4.95073 & -1.1009 \\
			0.215 & 3.67449 & 4.97533 & -1.12689 \\
			\hline 
		\end{tabular} \>\begin{tabular}{|p{0.8cm}|p{0.8cm}|p{0.9cm}p{1.3cm}|}	
		\multicolumn{4}{|c|}{$d=11$; $P_{c}=0.366593$ and $T=0.647545$}\\
		\hline
		\centering{$P$} & \centering{$r_{H}$} & \centering{$\omega_{r}$}  & \hspace{9pt}$\omega_{im}$\\ 
		\hline
		0.368 & 1.11035 & 4.55202 & -0.149345 \\
		0.3676 & 1.11051 & 4.54943 & -0.149007 \\
		0.3672 & 1.11066 & 4.54683 & -0.148671 \\
		0.367 & 1.11074 & 4.54554 & -0.148502 \\
		0.3668 & 1.11081 & 4.54424 & -0.148334 \\
		0.3666 & 1.11089 & 4.54294 & -0.148166 \\
		\hline\hline
		0.365 & 2.23893 & 5.30884 & -0.758411 \\
		0.36 & 2.36416 & 5.3839 & -0.814212 \\
		0.355 & 2.47173 & 5.44066 & -0.858141 \\
		0.35 & 2.57093 & 5.48771 & -0.895714 \\
		0.34 & 2.75731 & 5.56482 & -0.959642 \\
		0.33 & 2.93736 & 5.62801 & -1.01428 \\
		\hline 
	\end{tabular}
\end{tabbing}
\caption{The QNM frequencies of the massless scalar perturbation for black holes with different dimensions and sizes in the isothermal process.}\label{tab2}
\end{minipage}
\end{center}
\end{table}

\begin{figure}[!ht]
	\hspace*{-0.8cm}\begin{minipage}{1.13\linewidth}
		\begin{tabbing}
			\hspace{4.5cm}\=\hspace{4.5cm}\=\hspace{4.5cm}\=\kill
			\includegraphics[width=4.5cm,height=3.cm]{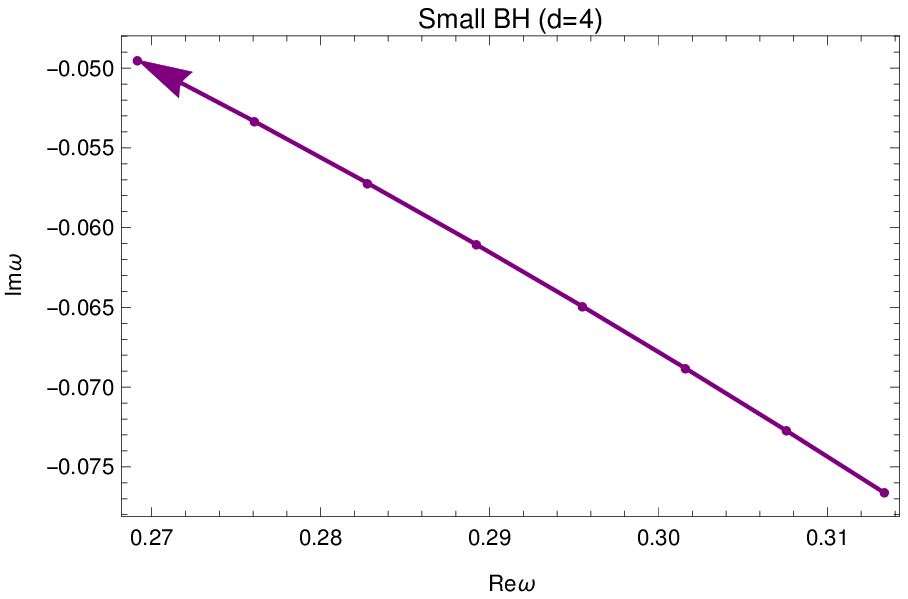}\> \includegraphics[width=4.5cm,height=3.cm]{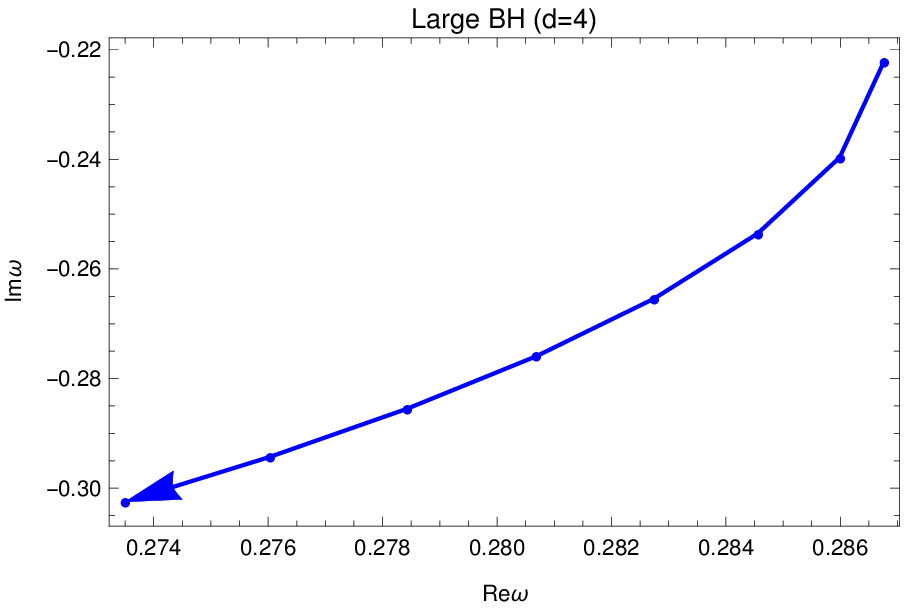} \>
			\includegraphics[width=4.5cm,height=3.cm]{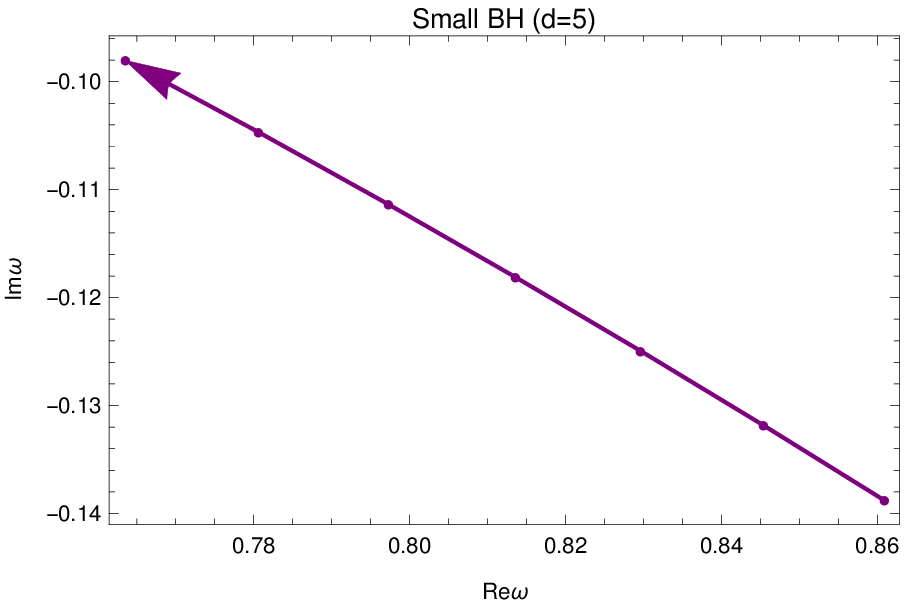}\> \includegraphics[width=4.5cm,height=3.cm]{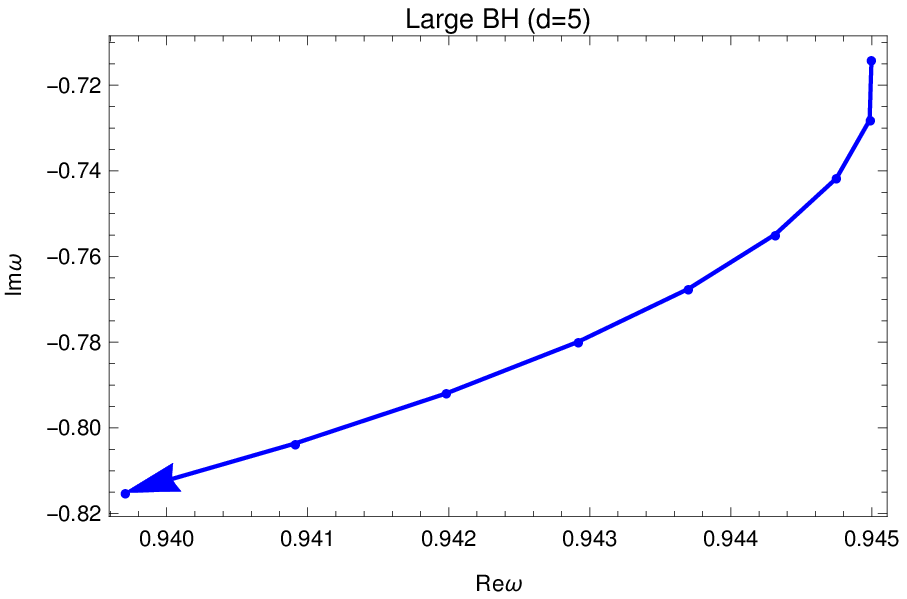} \\
			\includegraphics[width=4.5cm,height=3.cm]{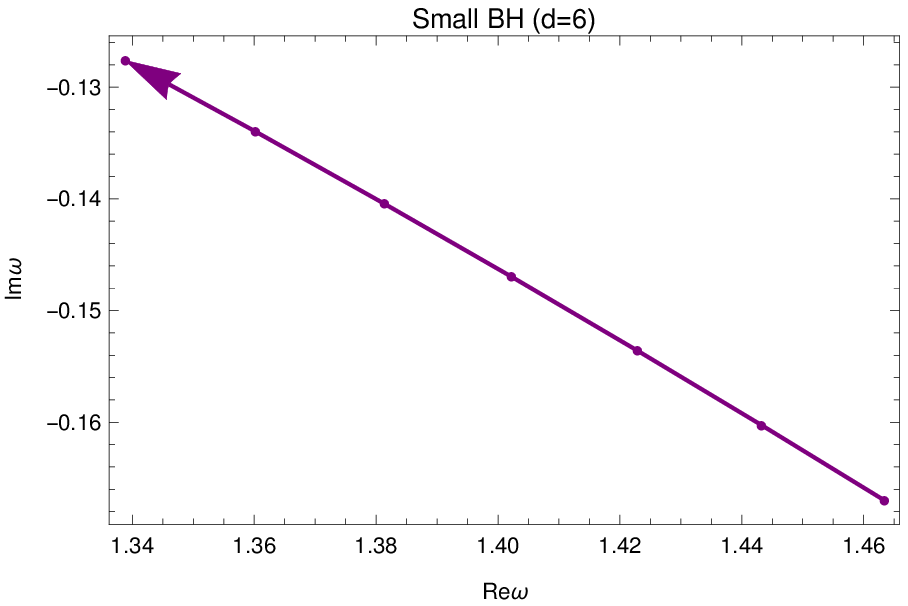}\> \includegraphics[width=4.5cm,height=3.cm]{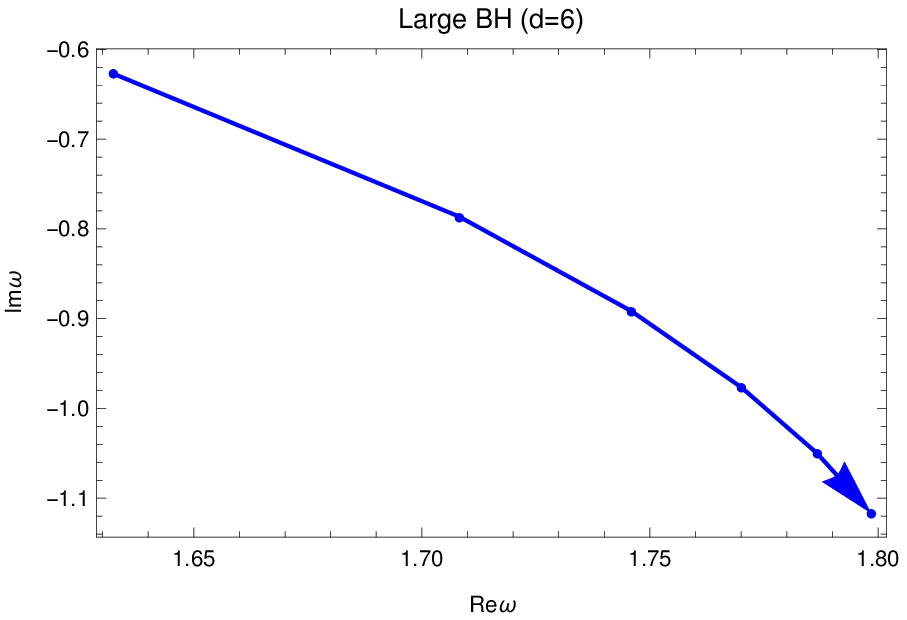} \>
			\includegraphics[width=4.5cm,height=3.cm]{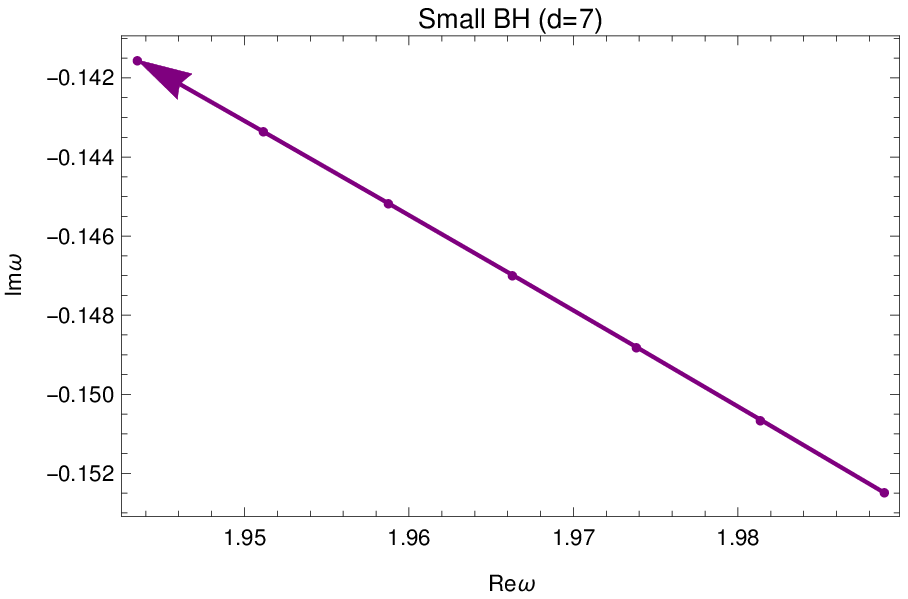}\> \includegraphics[width=4.5cm,height=3.cm]{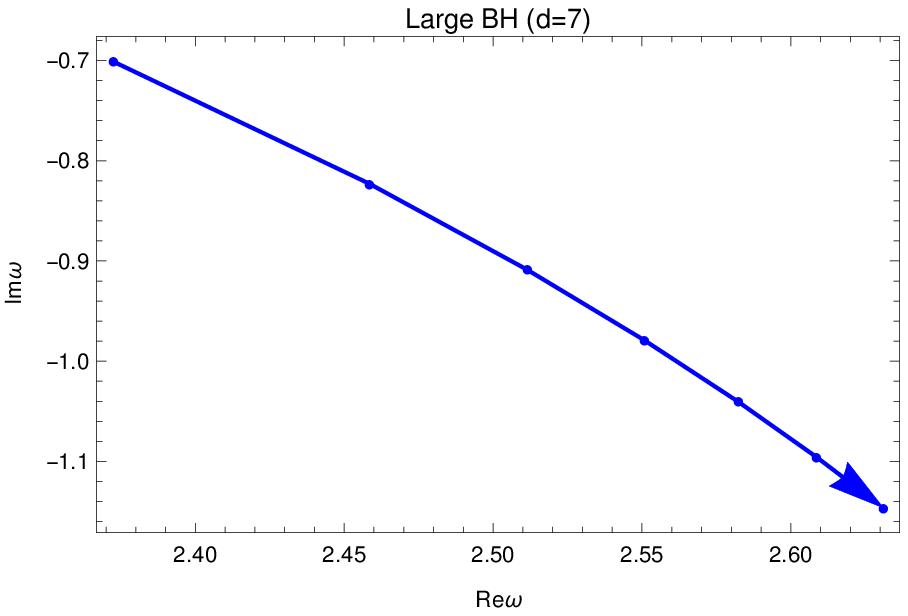} \\
			\includegraphics[width=4.5cm,height=3.cm]{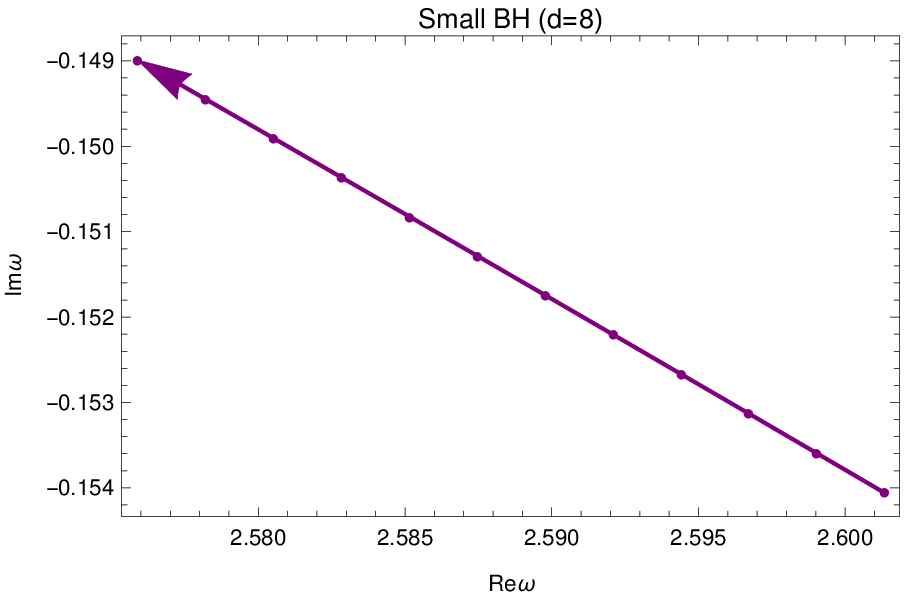}\> \includegraphics[width=4.5cm,height=3.cm]{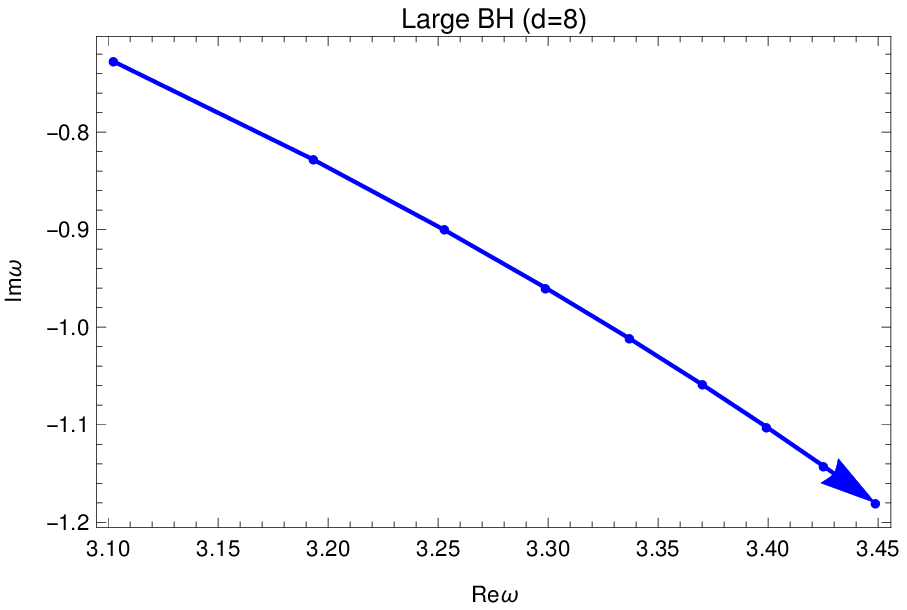} \>
			\includegraphics[width=4.5cm,height=3.cm]{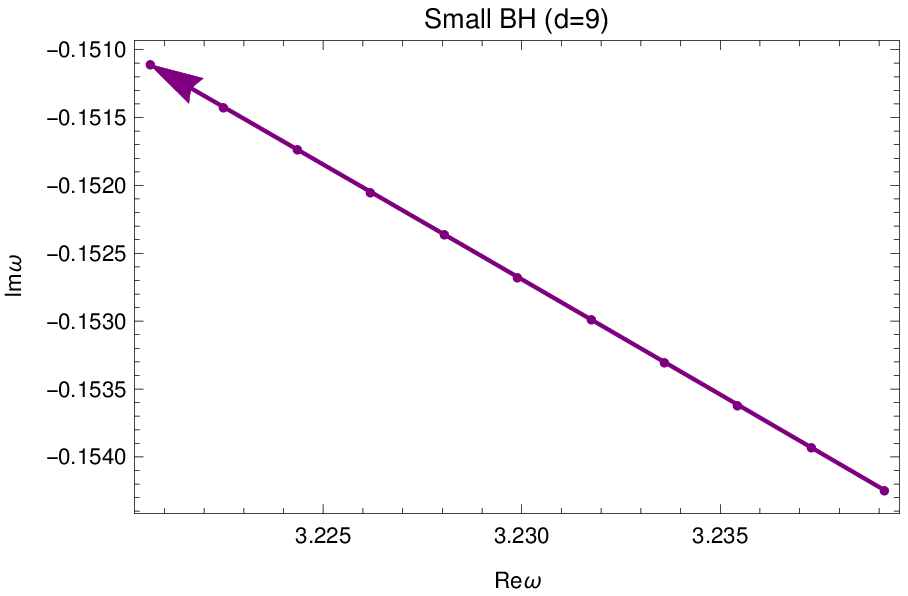}\> \includegraphics[width=4.5cm,height=3.cm]{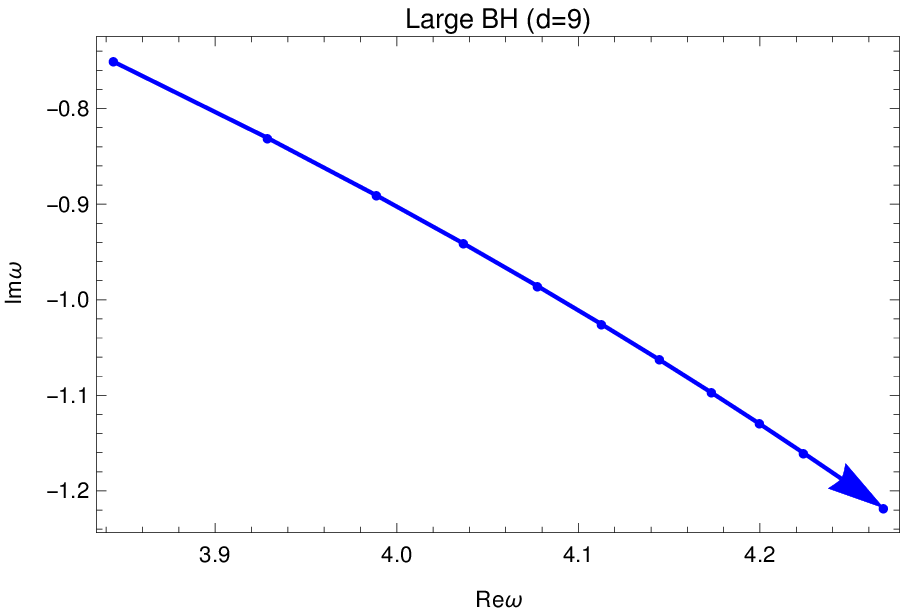} \\
			\includegraphics[width=4.5cm,height=3.cm]{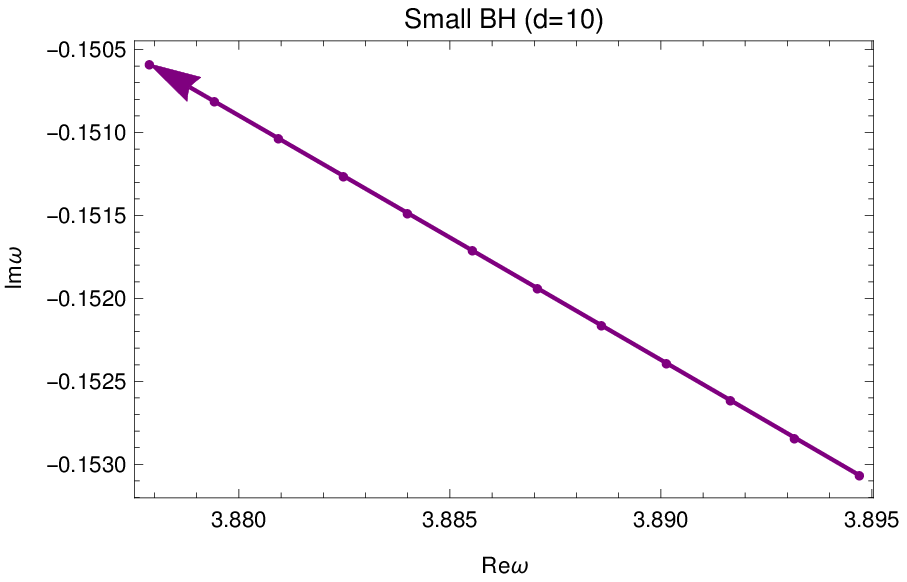}\> \includegraphics[width=4.5cm,height=3.cm]{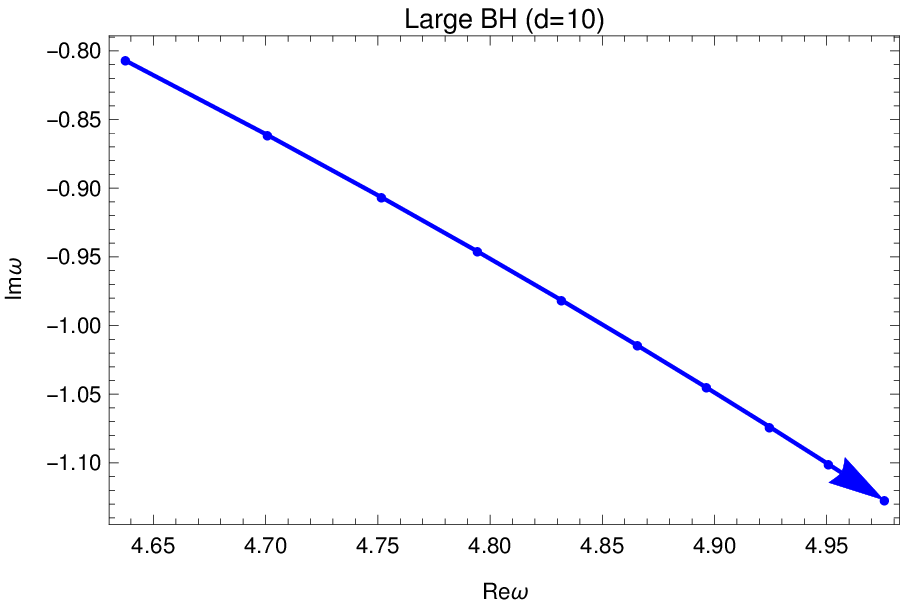} \>
			\includegraphics[width=4.5cm,height=3.cm]{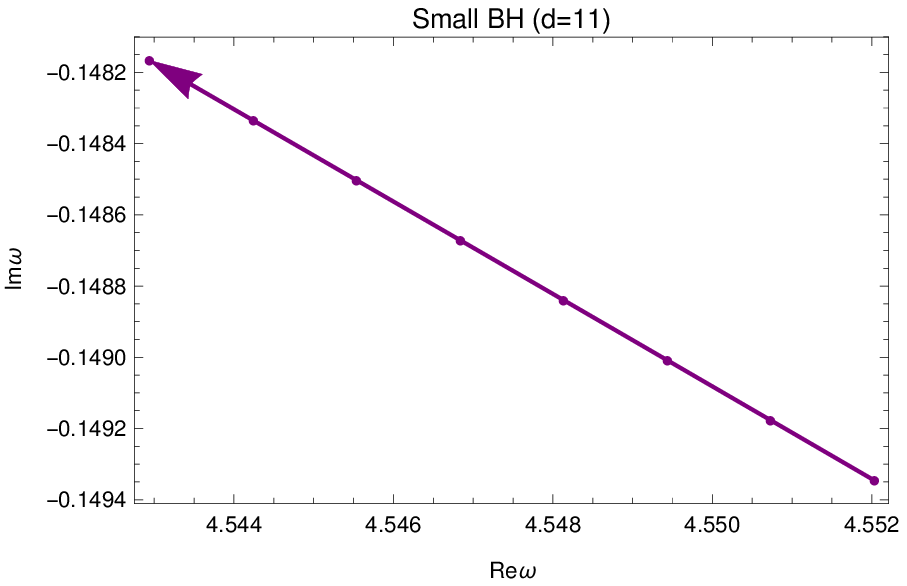}\> \includegraphics[width=4.5cm,height=3.cm]{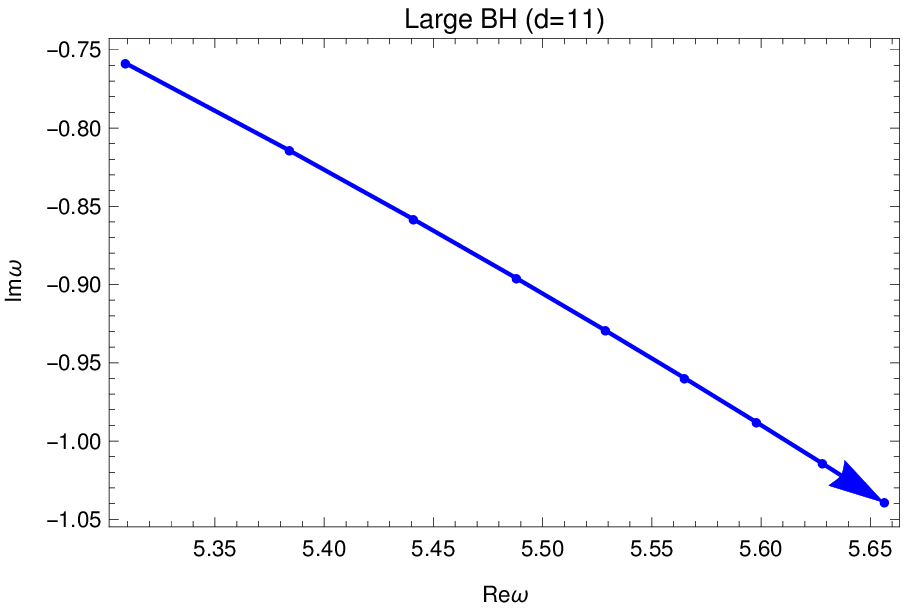} \\
		\end{tabbing}
		\vspace*{0cm} \caption{The behavior of the quasinormal modes for small and large black holes for a $4\leq d\leq 11$. 
			Increase in the black hole size is shown by arrows in the case of the isothermal process.}
		\label{fig5}
	\end{minipage}
\end{figure}

From  Figure \ref{fig5}, we can see that the behavior of the quasinormal modes in the isothermal process is not  regular and depends on the spacetime dimension. Indeed: 
\begin{itemize}
\item For $d=4$ and $d=5$, we see that for the SBH phase, the corresponding pressure $P$ decreases with growing black hole horizon. In this process, the real part of the frequencies as well as the absolute imaginary part decrease. For the LBH case, the pressure also decreases when the black hole size increases. However, the real  and the absolute values of the imaginary parts have opposite behaviors. Figure \ref{fig5} shows the behaviors of the quasinormal modes for small and large holes where the arrows indicate increase of the black hole size. It is clear that the remarkable properties of the  quasinormal frequencies are completely different in SBH and LBH phases, confirming behavior seen in $d=4$ \cite{base}, and providing a good measure to probe the black hole phase transitions.
\item For $d \ge 6$, the behaviors of QNM in the two  black hole phases are completely different: when the horizon radius grows the real and absolute value of the imaginary parts both decrease (increase) in the SBH phase (LBH phase), respectively. Therefore the QNM behavior is sensitive to the dimension $d$ which  affects the correction induced by variation of the horizon radius $r_H$ and the pressure $P$, as can be seen from the master equation Eq. \eqref{psieqn01}.
 \end{itemize}
 
\begin{figure}[!ht]
		\hspace*{-0.8cm}\begin{minipage}{1.13\linewidth}
			\begin{tabbing}
				\hspace{4.5cm}\=\hspace{4.5cm}\=\hspace{4.5cm}\=\kill
				\includegraphics[width=4.5cm,height=3.cm]{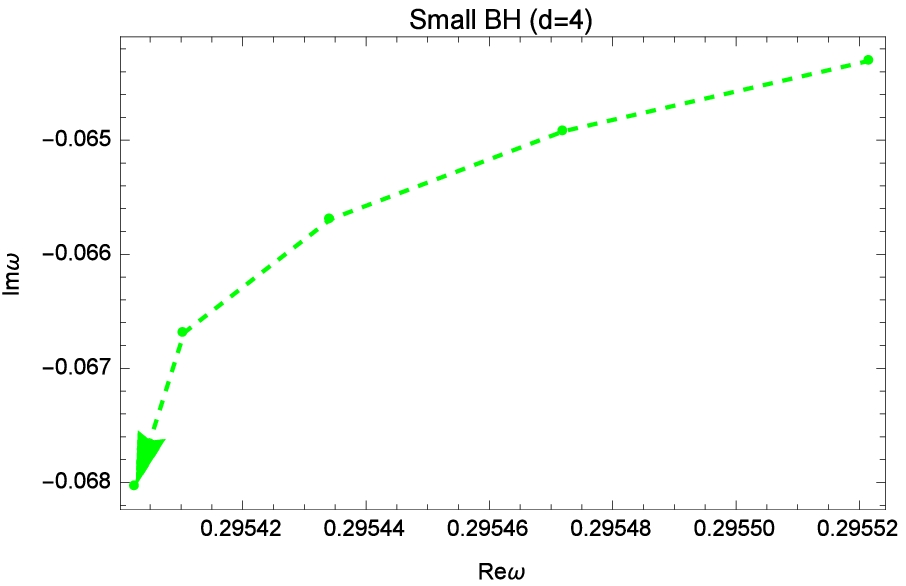}\> \includegraphics[width=4.5cm,height=3.cm]{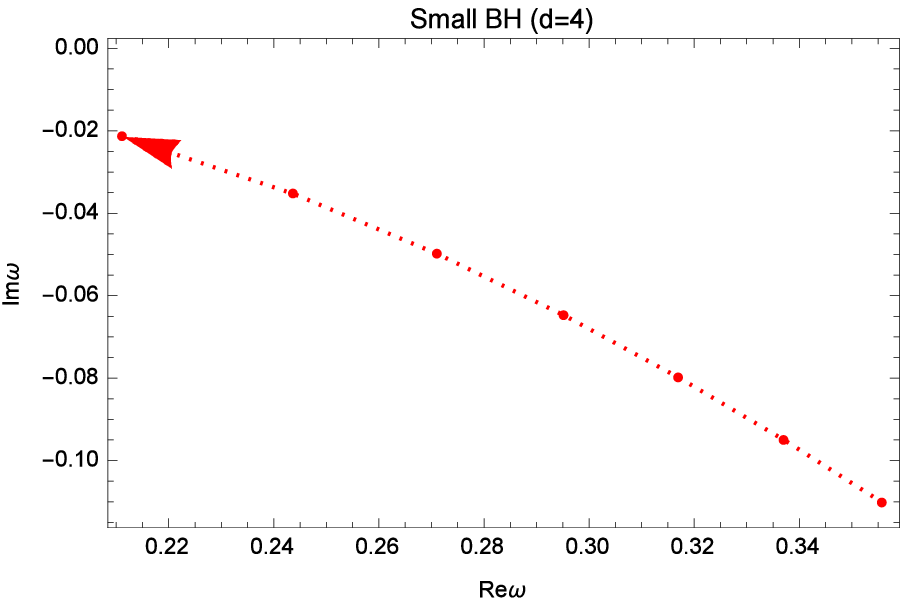} \>
				\includegraphics[width=4.5cm,height=3.cm]{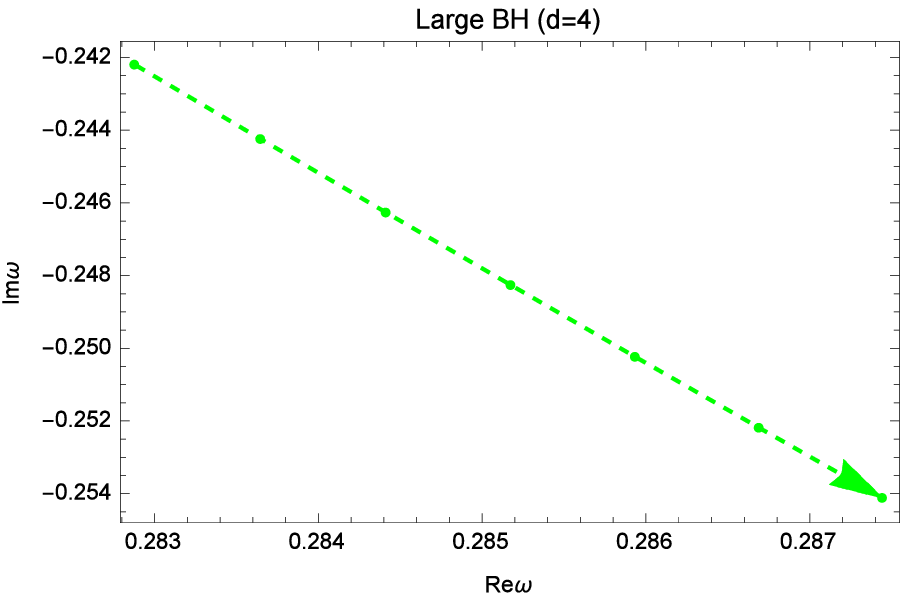}\> \includegraphics[width=4.5cm,height=3.cm]{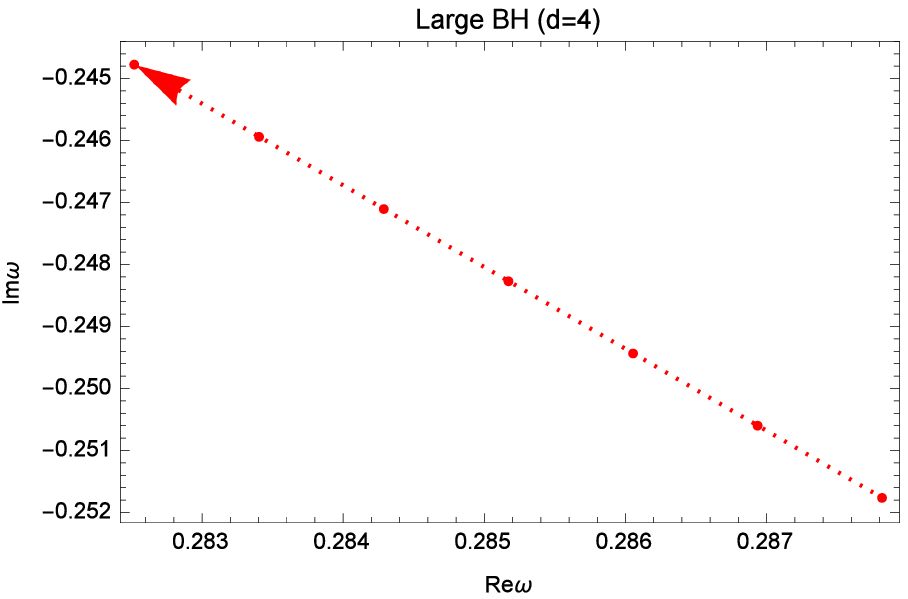} \\
				\includegraphics[width=4.5cm,height=3.cm]{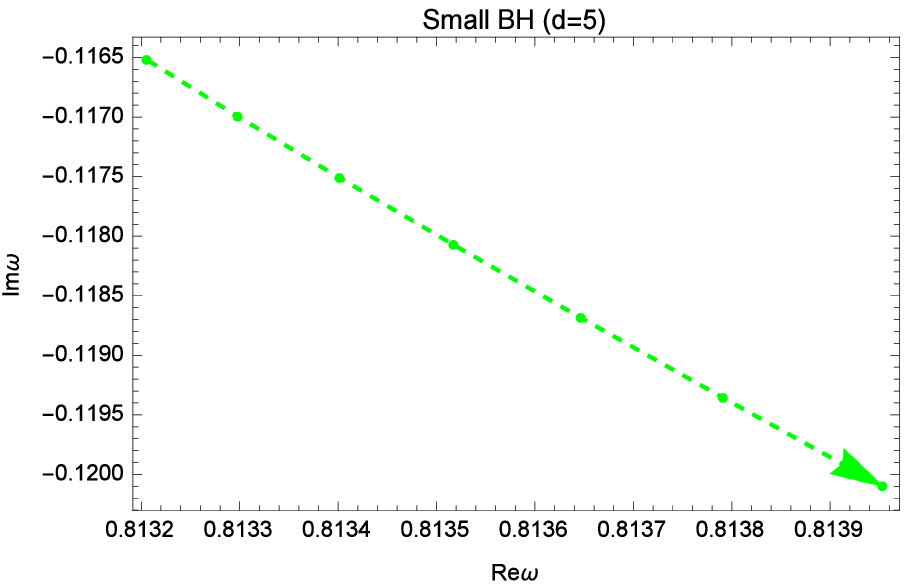}\> \includegraphics[width=4.5cm,height=3.cm]{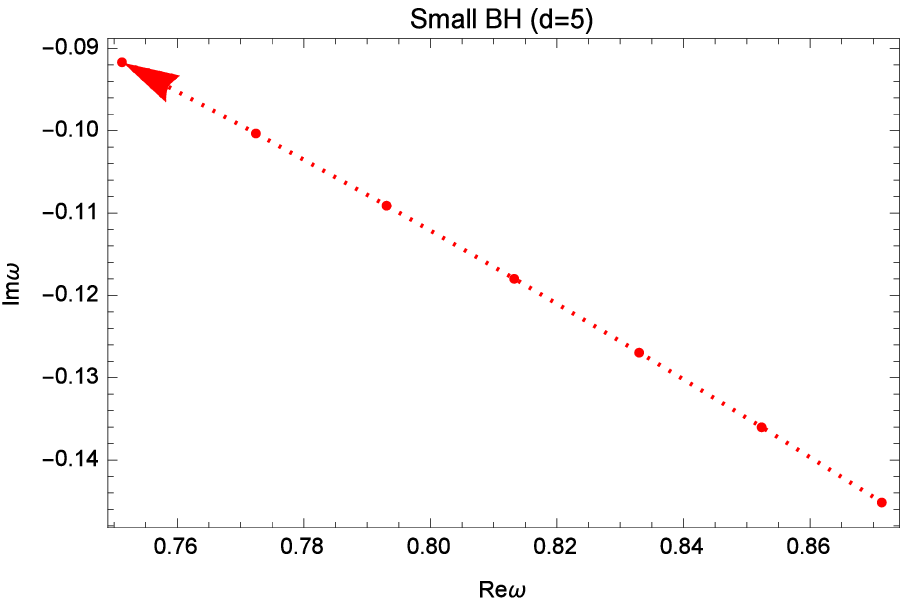} \>
				\includegraphics[width=4.5cm,height=3.cm]{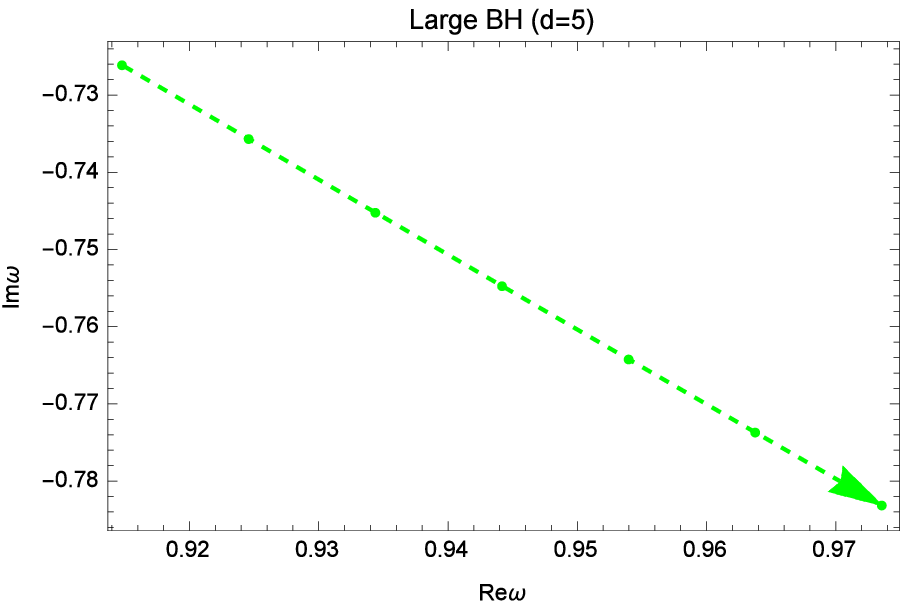}\> \includegraphics[width=4.5cm,height=3.cm]{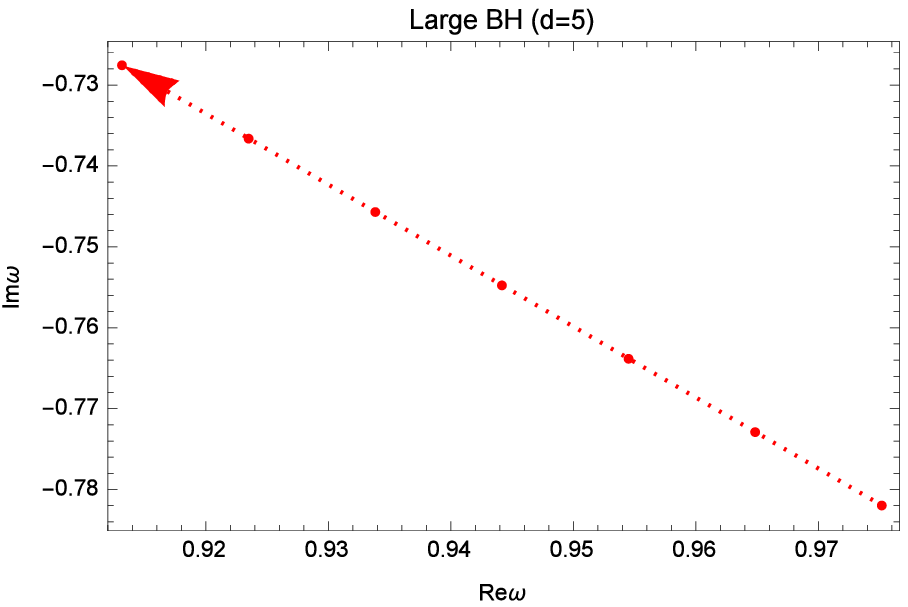} \\
				\includegraphics[width=4.5cm,height=3.cm]{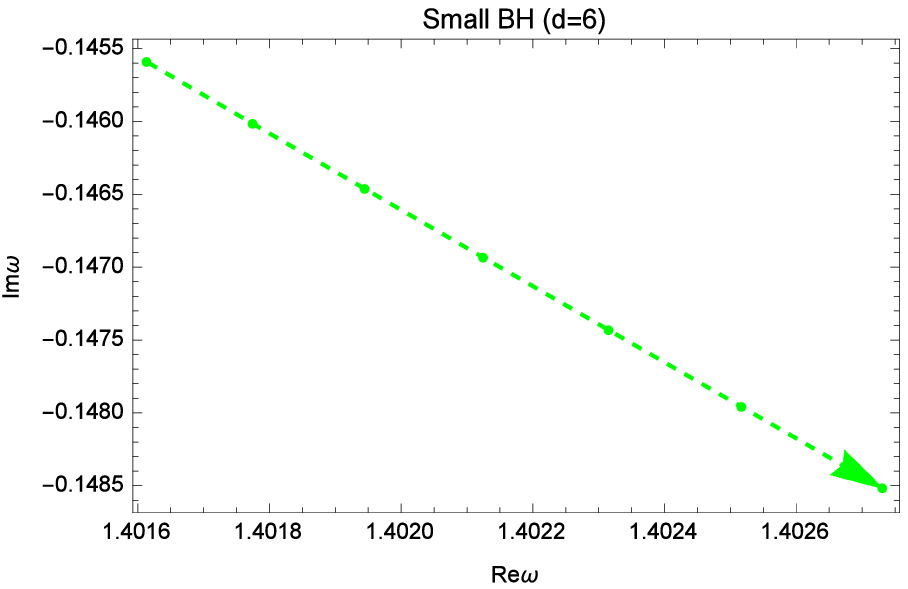}\> \includegraphics[width=4.5cm,height=3.cm]{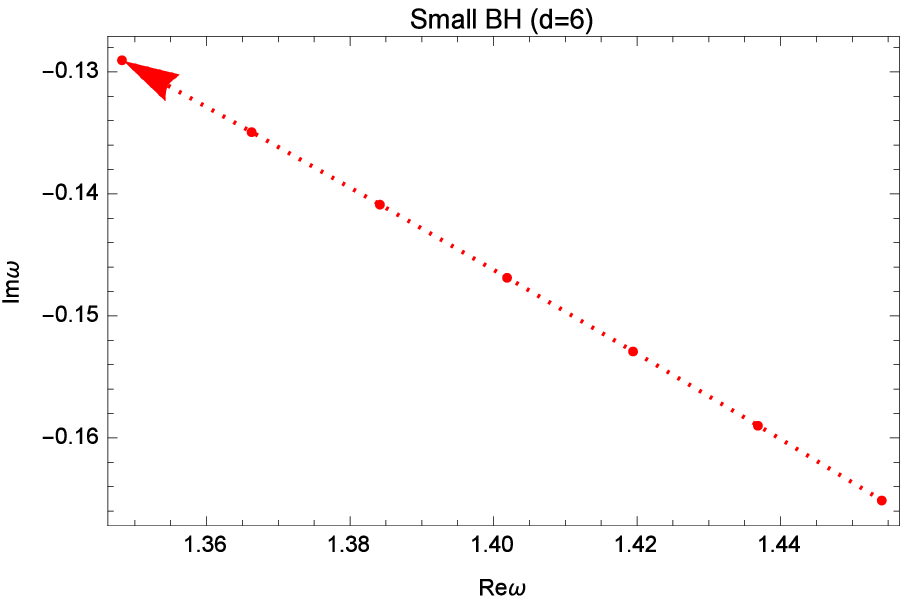} \>
				\includegraphics[width=4.5cm,height=3.cm]{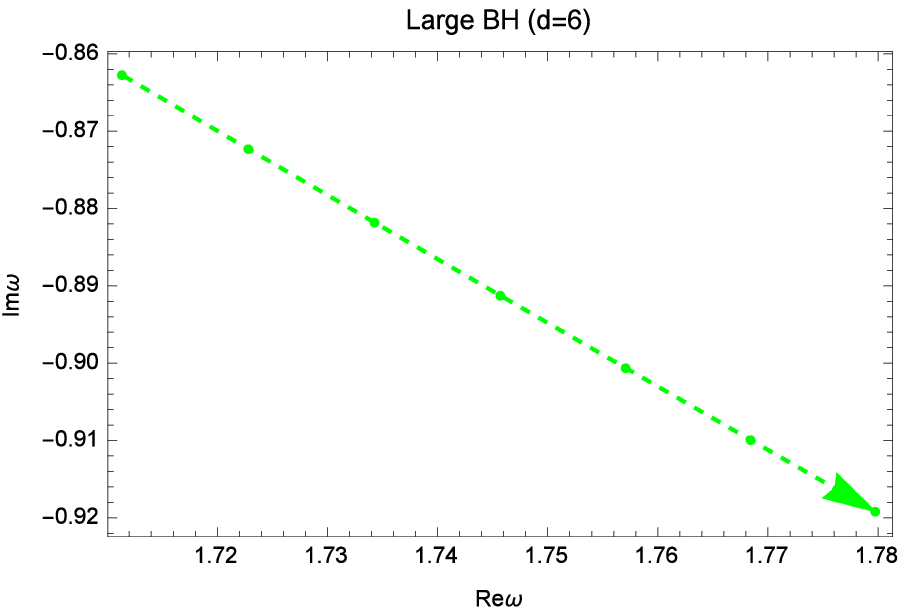}\> \includegraphics[width=4.5cm,height=3.cm]{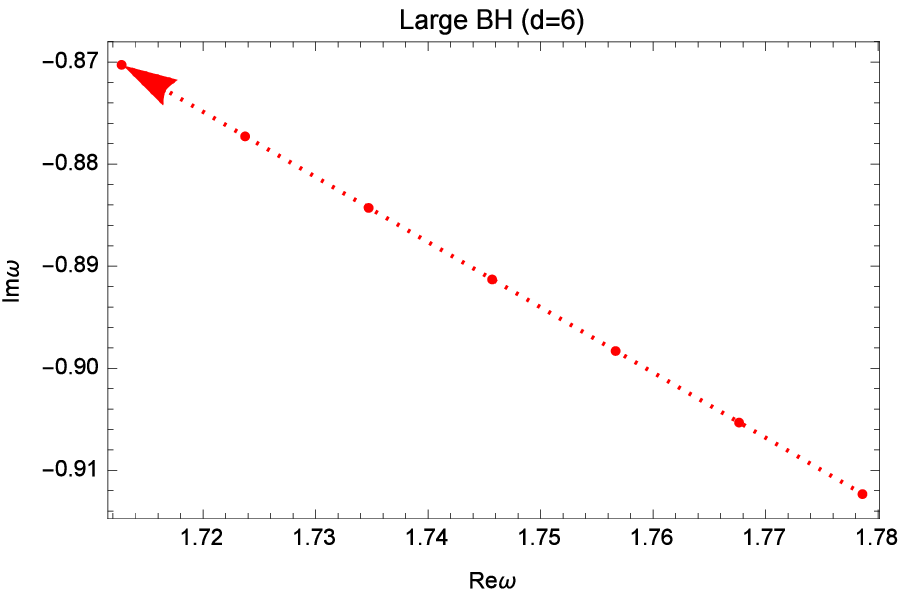} \\
			\end{tabbing}
			\vspace*{0cm} \caption{The purple dashed plots shows the quasinormal frequencies variation  with the horizon  $r_H$, for fixed pressure $P$. The blue dotted represents the quasinormal frequencies as a function of $P$, for fixed black hole horizons $r_H$ (Table \ref{ta3}). The arrow indicates the increase of $r_H$ ($P$) for fixed $P$ ($r_H$), respectively.}\label{fig6}
		\end{minipage}
	\end{figure}

 In the isothermal transition, the quasinormal modes can be affected by the value of the pressure $P(l)$ and the horizon radius $r_H$. To illustrate  the effect of  these two parameters, we restrict our  calculation here to only  $d=4$, $d=5$ and $d=6$. For higher dimensions, $d>6$, the behavior is exactly similar to $d=6$ as shown in Table \ref{ta3}. This feature can also be seen clearly in Figure \ref{fig6} where we notice a competition between the pressure $P$ and horizon radius $r_H$. Each of these parameters aims to overwhelm the other which affects the decay rate of the field, namely $\omega_i$. 

\begin{table}[!ht]\tiny
	\begin{center}
		\begin{minipage}{0.88\linewidth}
			\begin{tabbing}
				\hspace{8.4cm}\=\kill
				\begin{tabular}{|p{0.8cm}|p{0.8cm}|p{0.9cm}p{1.3cm}|}	
					\multicolumn{4}{|c|}{Small BH ($d=4$)}\\	
					\hline
					\centering{$P$} & \centering{$r_{H}$} & \centering{$\omega_{r}$}  & \hspace{9pt}$\omega_{im}$\\ 
					\hline
					& 1.32238 & 0.295522 & -0.0643022 \\
					& 1.35161 & 0.295472 & -0.0649205 \\
					0.002 & 1.38539 & 0.295434 & -0.065692 \\
					& 1.42549 & 0.295411 & -0.0666873 \\
					& 1.47507 & 0.295403 & -0.0680322 \\
					\hline 
				\end{tabular}
				\>\begin{tabular}{|p{0.8cm}|p{0.8cm}|p{0.9cm}p{1.3cm}|}
					\multicolumn{4}{|c|}{Small BH ($d=4$)}\\
					\hline
					\centering{$r_{H}$} & \centering{$P$} & \centering{$\omega_{r}$}  & \hspace{9pt}$\omega_{im}$\\ 
					\hline
					& 0.00201306 & 0.296278 & -0.0654519 \\
					& 0.00200653 & 0.295875 & -0.0651862 \\
					1.352 & 0.002 & 0.295472 & -0.0649205 \\
					& 0.00199347 & 0.295068 & -0.0646549 \\
					& 0.00198694 & 0.294664 & -0.0643894 \\
					\hline 
				\end{tabular} \\ \\
				\begin{tabular}{|p{0.8cm}|p{0.8cm}|p{0.9cm}p{1.3cm}|}	
					\multicolumn{4}{|c|}{Large BH ($d=4$)}\\	
					\hline
					\centering{$P$} & \centering{$r_{H}$} & \centering{$\omega_{r}$}  & \hspace{9pt}$\omega_{im}$\\ 
					\hline
					& 7.78691 & 0.283655 & -0.244264 \\
					& 7.85092 & 0.284421 & -0.246286 \\
					0.0014 & 7.91417 & 0.285183 & -0.248283 \\
					& 7.9767 & 0.285942 & -0.250256 \\
					& 8.03854 & 0.286699 & -0.252208 \\
					\hline 
				\end{tabular}\>\begin{tabular}{|p{0.8cm}|p{0.8cm}|p{0.9cm}p{1.3cm}|}	
				\multicolumn{4}{|c|}{Large BH ($d=4$)}\\
				\hline
				\centering{$r_{H}$} & \centering{$P$} & \centering{$\omega_{r}$}  & \hspace{9pt}$\omega_{im}$\\ 
				\hline
				& 0.00141306 & 0.286948 & -0.250613 \\
				& 0.00140653 & 0.286066 & -0.249448 \\
				7.914 & 0.0014 & 0.285183 & -0.248283 \\
				& 0.00139347 & 0.2843 & -0.247118 \\
				& 0.00138694 & 0.283416 & -0.245953 \\
				\hline 
			\end{tabular}\\ \\
			\begin{tabular}{|p{0.8cm}|p{0.8cm}|p{0.9cm}p{1.3cm}|}	
				\multicolumn{4}{|c|}{Small BH ($d=5$)}\\
				\hline
				\centering{$P$} & \centering{$r_{H}$} & \centering{$\omega_{r}$}  & \hspace{9pt}$\omega_{im}$\\ 
				\hline
				& 1.23887 & 0.8133 & -0.117003 \\
				& 1.24662 & 0.813403 & -0.117518 \\
				0.012 & 1.2548 & 0.813519 & -0.118079 \\
				& 1.26348 & 0.813648 & -0.118692 \\
				& 1.2727 & 0.813793 & -0.119364 \\
				\hline 
			\end{tabular} \> \begin{tabular}{|p{0.8cm}|p{0.8cm}|p{0.9cm}p{1.3cm}|}	
			\multicolumn{4}{|c|}{Small BH ($d=5$)}\\
			\hline
			\centering{$r_{H}$} & \centering{$P$} & \centering{$\omega_{r}$}  & \hspace{9pt}$\omega_{im}$\\ 
			\hline
			& 0.0132254 & 0.852622 & -0.136113 \\
			& 0.0126127 & 0.833267 & -0.127053 \\
			1.255 & 0.012 & 0.813519 & -0.118079 \\
			& 0.0113873 & 0.793341 & -0.109201 \\
			& 0.0107746 & 0.772694 & -0.100431 \\
			\hline 
		\end{tabular}\\ \\
		\begin{tabular}{|p{0.8cm}|p{0.8cm}|p{0.9cm}p{1.3cm}|}	
			\multicolumn{4}{|c|}{Large BH ($d=5$)}\\
			\hline
			\centering{$P$} & \centering{$r_{H}$} & \centering{$\omega_{r}$}  & \hspace{9pt}$\omega_{im}$\\ 
			\hline
			& 21.5956 & 0.924718 & -0.735804 \\
			& 21.8686 & 0.934517 & -0.745342 \\
			0.003 & 22.141 & 0.944314 & -0.754857 \\
			& 22.4128 & 0.954109 & -0.764347 \\
			& 22.684 & 0.963901 & -0.773816 \\
			\hline 
		\end{tabular} \> \begin{tabular}{|p{0.8cm}|p{0.8cm}|p{0.9cm}p{1.3cm}|}	
		\multicolumn{4}{|c|}{Large BH ($d=5$)}\\
		\hline
		\centering{$r_{H}$} & \centering{$P$} & \centering{$\omega_{r}$}  & \hspace{9pt}$\omega_{im}$\\ 
		\hline
		& 0.00307123 & 0.964986 & -0.772998 \\
		& 0.00303561 & 0.95465 & -0.763927 \\
		22.141 & 0.003 & 0.944314 & -0.754857 \\
		& 0.00296439 & 0.933977 & -0.745786 \\
		& 0.00292877 & 0.923639 & -0.736716 \\
		\hline 
	\end{tabular}\\ \\
	
	\begin{tabular}{|p{0.8cm}|p{0.8cm}|p{0.9cm}p{1.3cm}|}	
		\multicolumn{4}{|c|}{Small BH ($d=6$)}\\
		\hline
		\centering{$P$} & \centering{$r_{H}$} & \centering{$\omega_{r}$}  & \hspace{9pt}$\omega_{im}$\\ 
		\hline
		& 1.19496 & 1.40178 & -0.146022 \\
		& 1.19874 & 1.40195 & -0.146469 \\
		0.034 & 1.20266 & 1.40213 & -0.14694 \\
		& 1.20672 & 1.40232 & -0.147438 \\
		& 1.21094 & 1.40252 & -0.147965 \\
		\hline 
	\end{tabular} \> \begin{tabular}{|p{0.8cm}|p{0.8cm}|p{0.9cm}p{1.3cm}|}	
	\multicolumn{4}{|c|}{Small BH ($d=6$)}\\
	\hline
	\centering{$r_{H}$} & \centering{$P$} & \centering{$\omega_{r}$}  & \hspace{9pt}$\omega_{im}$\\ 
	\hline
	& 0.0356841 & 1.43709 & -0.159068 \\
	& 0.034842 & 1.41969 & -0.152981 \\
	1.203 & 0.034 & 1.40213 & -0.14694 \\
	& 0.033158 & 1.38441 & -0.140946 \\
	& 0.0323159 & 1.36652 & -0.135001 \\
	\hline 
\end{tabular}\\ \\
\begin{tabular}{|p{0.8cm}|p{0.8cm}|p{0.9cm}p{1.3cm}|}	
	\multicolumn{4}{|c|}{Large BH ($d=6$)}\\
	\hline
	\centering{$P$} & \centering{$r_{H}$} & \centering{$\omega_{r}$}  & \hspace{9pt}$\omega_{im}$\\ 
	\hline
	& 6.84613 & 1.72298 & -0.872432 \\
	& 6.91299 & 1.73444 & -0.881938 \\
	0.02 & 6.97943 & 1.74585 & -0.891377 \\
	& 7.04545 & 1.75723 & -0.900751 \\
	& 7.11109 & 1.76858 & -0.910063 \\
	\hline 
\end{tabular} \> \begin{tabular}{|p{0.8cm}|p{0.8cm}|p{0.9cm}p{1.3cm}|}	
\multicolumn{4}{|c|}{Large BH ($d=6$)}\\
\hline
\centering{$r_{H}$} & \centering{$P$} & \centering{$\omega_{r}$}  & \hspace{9pt}$\omega_{im}$\\ 
\hline
& 0.0202979 & 1.7678 & -0.905398 \\
& 0.0201489 & 1.75683 & -0.898387 \\
6.979 & 0.02 & 1.74585 & -0.891377 \\
& 0.0198511 & 1.73488 & -0.884367 \\
& 0.0197021 & 1.7239 & -0.877357 \\
\hline 
\end{tabular}
\end{tabbing}
\caption{Left: the quasinormal frequencies as function of the black hole horizon  $r_H$, for fixed pressure $P$. Right: quasinormal frequencies as function of $P$, for fixed black hole horizons $r_H$.}\label{ta3}
\end{minipage}
\end{center}
\end{table} 

To illustrate how these two factors affect the quasinormal frequencies, we  perform a double-series expansion of the frequency $\omega\left(r_{H}+\Delta r_{H},P+\Delta P\right)$ \cite{base}:
\begin{equation}
\omega\left(r_{H}+\Delta r_{H},P+\Delta P\right)=\omega\left(r_{H},P \right)+\frac{\partial \omega}{\partial r_{H}}\Delta r_{H}+\frac{\partial \omega}{\partial P}\Delta P+ \mathcal{O}(\Delta r_{H}^{2},\Delta P^{2},\Delta r_{H}.\Delta P),
\end{equation}

with, 
$$\tilde{\omega}=\omega+\Delta_{r_{H}}+\Delta_{p}+ \mathcal{O}(\Delta r_{H}^{2},\Delta P^{2},\Delta r_{H}.\Delta P),$$
where $\Delta_{r_{H}}$ and $\Delta_{p}$ represent corrections induced by variation of the black hole size and pressure, respectively.  Recall here that the choice of the step of pressure $\Delta P$ in linear approximation is related to $\Delta r_H$ via the relation:
\begin{equation}\label{dpeqt}
dP=\left(-\frac{(d-2) T}{4 r_H^2}+\frac{(3-d) (d-2)^2 r_H^{(3-2 d)}}{8 \pi }+\frac{(d-3) (d-2)}{8 \pi  r_H^3}\right) dr_H,
\end{equation}
Table \ref{ta4} shows the evaluation of the quasinormal frequencies $\tilde{\omega}$ from the linear approximation for
small and large black hole phase\footnote{This evaluation is performed in a narrow range of $r_H$ in order to preserve the linear approximation in Eq. \eqref{dpeqt}}. One can see that the behaviors of $\tilde{\omega}$ are consistent with the numerical computation. For exemple for small black hole at $d=5$ we obtain $\tilde{\omega_{r}}<\omega_{r}$ and $|\tilde{\omega}_{im}|<|\omega_{im}|$, which confirms the behavior shown in Figure \ref{fig5}.
\begin{table}[!ht]\tiny
	\begin{center}
		\hspace*{-0.8cm}\begin{minipage}{1.12\linewidth}
			\begin{tabbing}
				
				{\footnotesize
					\begin{tabular}{|p{0.9cm}|p{0.8cm}|p{3cm}|p{3cm}|p{3.5cm}|p{3.9cm}|}	
						\multicolumn{6}{|c|}{small BH ($d=4$)}\\
					\hline
						\centering{$P$} & \centering{$r_{H}$} & \centering{$\omega$}  & \centering{$\tilde{\omega}$} & \centering{$\Delta_{p}$} & \hspace{1.8cm}$\Delta_{r_{H}}$ \\ 
						\hline
						0.002 & 1.352 & 0.295472\, -0.0649205 i & 0.289516\, -0.0611853 i & -0.00594507+0.00391361 i & -0.0000114725-0.000178387 i \\
						0.0019 & 1.360 & 0.289191\, -0.0610417 i & 0.283101\, -0.05734 i & -0.00607645+0.00388441 i & -0.0000133426-0.00018272 i \\
						0.0017 & 1.379 & 0.276049\, -0.0533207 i & 0.269655\, -0.0496962 i & -0.00637566+0.00381701 i & -0.0000186237-0.000192563 i \\
						\hline 
					\end{tabular}
				} \\ \\
				{\footnotesize
					\begin{tabular}{|p{0.9cm}|p{0.8cm}|p{3cm}|p{3cm}|p{3.5cm}|p{3.9cm}|}	
						\multicolumn{6}{|c|}{large BH ($d=4$)}\\
						\hline
						\centering{$P$} & \centering{$r_{H}$} & \centering{$\omega$}  & \centering{$\tilde{\omega}$} & \centering{$\Delta_{p}$} & \hspace{1.8cm}$\Delta_{r_{H}}$ \\ 
						\hline
						0.0014 & 7.914 & 0.285183\, -0.248283 i & 0.284939\, -0.250513 i & -0.00222237+0.00293185 i & 0.00197779\, -0.005162 i \\
						0.00135 & 8.422 & 0.284389\, -0.254799 i & 0.284128\, -0.256734 i & -0.00219918+0.00292837 i & 0.00193895\, -0.00486306 i \\
						0.0013 & 8.952 & 0.283505\, -0.260848 i & 0.283237\, -0.262547 i & -0.00217896+0.00292535 i & 0.00191109\, -0.00462342 i \\
						\hline 
					\end{tabular}
				}\\ \\
				{\footnotesize
					\begin{tabular}{|p{0.9cm}|p{0.8cm}|p{3cm}|p{3cm}|p{3.5cm}|p{3.9cm}|}
						\multicolumn{6}{|c|}{small BH ($d=5$)}\\
						\hline
						\centering{$P$} & \centering{$r_{H}$} & \centering{$\omega$}  & \centering{$\tilde{\omega}$} & \centering{$\Delta_{p}$} & \hspace{1.8cm}$\Delta_{r_{H}}$ \\  
						\hline
						0.013 & 1.241 & 0.845309\, -0.131794 i & 0.840402\, -0.129622 i & -0.00494032+0.00231399 i & 0.0000336195\, -0.000141584 i \\
						0.012 & 1.255 & 0.813519\, -0.118079 i & 0.808489\, -0.115967 i & -0.00506175+0.0022644 i & 0.000031792\, -0.000152273 i \\
						0.011 & 1.270 & 0.780515\, -0.104628 i & 0.775344\, -0.102586 i & -0.00519991+0.00220809 i & 0.0000287101\, -0.000165456 i \\
						\hline 
					\end{tabular}
				}\\ \\
				{\footnotesize
					\begin{tabular}{|p{0.9cm}|p{0.8cm}|p{3cm}|p{3cm}|p{3.5cm}|p{3.9cm}|}
						\multicolumn{6}{|c|}{large BH ($d=5$)}\\	
						\hline
						\centering{$P$} & \centering{$r_{H}$} & \centering{$\omega$}  & \centering{$\tilde{\omega}$} & \centering{$\Delta_{p}$} & \hspace{1.8cm}$\Delta_{r_{H}}$ \\ 
						\hline
						0.004 & 16.100 & 0.944984\, -0.728163 i & 0.944981\, -0.728497 i & -0.00256351+0.00224264 i & 0.00256039\, -0.00257707 i \\
						0.003 & 22.141 & 0.944314\, -0.754857 i & 0.944304\, -0.755084 i & -0.00255623+0.00224307 i & 0.00254691\, -0.00247059 i \\
						0.001 & 70.112 & 0.940907\, -0.803661 i & 0.940901\, -0.803725 i & -0.00254884+0.00224348 i & 0.00254249\, -0.00230779 i \\
						\hline 
					\end{tabular}
				}\\ \\
				
				{\footnotesize
					\begin{tabular}{|p{0.9cm}|p{0.8cm}|p{3cm}|p{3cm}|p{3.5cm}|p{3.9cm}|}
						\multicolumn{6}{|c|}{small BH ($d=6$)}\\
						\hline
						\centering{$P$} & \centering{$r_{H}$} & \centering{$\omega$}  & \centering{$\tilde{\omega}$} & \centering{$\Delta_{p}$} & \hspace{1.8cm}$\Delta_{r_{H}}$ \\ 
						\hline
						0.034 & 1.203 & 1.40213\, -0.14694 i & 1.39765\, -0.145521 i & -0.0045282+0.00154515 i & 0.0000481149\, -0.000125901 i \\
						0.033 & 1.208 & 1.38128\, -0.140413 i & 1.37676\, -0.139015 i & -0.00456325+0.00152818 i & 0.0000477697\, -0.000130945 i \\
						0.032 & 1.213 & 1.36018\, -0.133963 i & 1.35562\, -0.132589 i & -0.00459992+0.00151046 i & 0.0000473116\, -0.000136514 i \\
						\hline 
					\end{tabular}
				}\\ \\
				{\footnotesize
					\begin{tabular}{|p{0.9cm}|p{0.8cm}|p{3cm}|p{3cm}|p{3.5cm}|p{3.9cm}|}	
						\multicolumn{6}{|c|}{large BH ($d=6$)}\\
						\hline
						\centering{$P$} & \centering{$r_{H}$} & \centering{$\omega$}  & \centering{$\tilde{\omega}$} & \centering{$\Delta_{p}$} & \hspace{1.8cm}$\Delta_{r_{H}}$ \\ 
						\hline
						0.025 & 5.067 & 1.70814\, -0.786255 i & 1.70865\, -0.787499 i & -0.00278508+0.00174631 i & 0.00328594\, -0.00299044 i \\
						0.02 & 6.979 & 1.74585\, -0.891377 i & 1.74607\, -0.892069 i & -0.00274466+0.0017533 i & 0.00296364\, -0.00244549 i \\
						0.01 & 15.876 & 1.78665\, -1.05007 i & 1.78669\, -1.05029 i & -0.00271047+0.00175882 i & 0.00275593\, -0.00198602 i \\
						\hline 
					\end{tabular}
				}
			\end{tabbing}
			\caption{$\tilde{\omega}$ is the quasinormal frequency from the linear approximation. $\Delta_{r_{H}}$ and $\Delta_{p}$ represent corrections due to variation of the black hole size and pressure, respectively}\label{ta4}
		\end{minipage}
	\end{center}
\end{table}

It is clear that for the small black hole case and for all dimensions, the change of the pressure prevails over the change of the black hole size. We also notice that the imaginary part of the frequency is more sensitive to the pressure.

\begin{figure}[!ht]
		\hspace*{-0.8cm}\begin{minipage}{1.13\linewidth}
			\begin{tabbing}
				\hspace{4.5cm}\=\hspace{4.5cm}\=\hspace{4.5cm}\=\kill
				\includegraphics[width=4.5cm,height=3.cm]{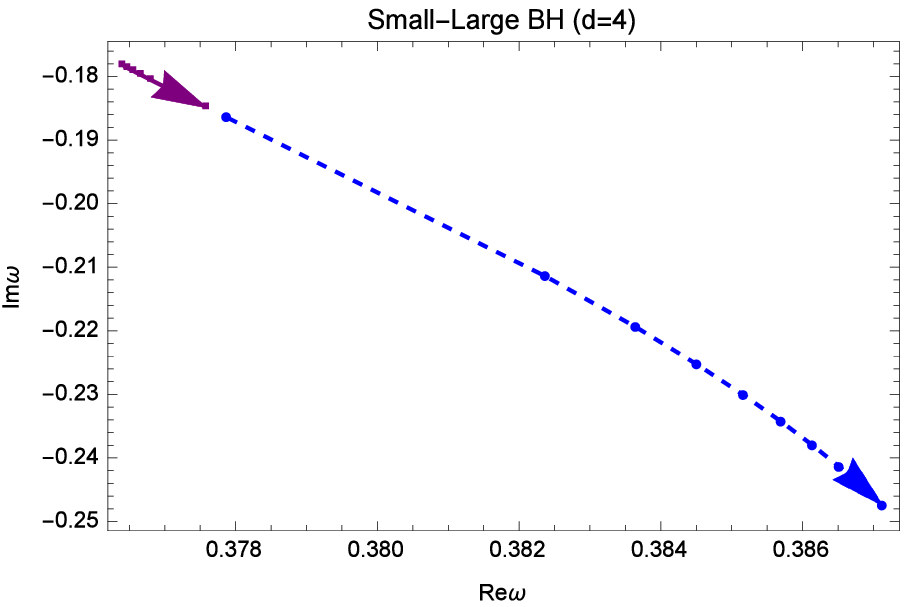}\> \includegraphics[width=4.5cm,height=3.cm]{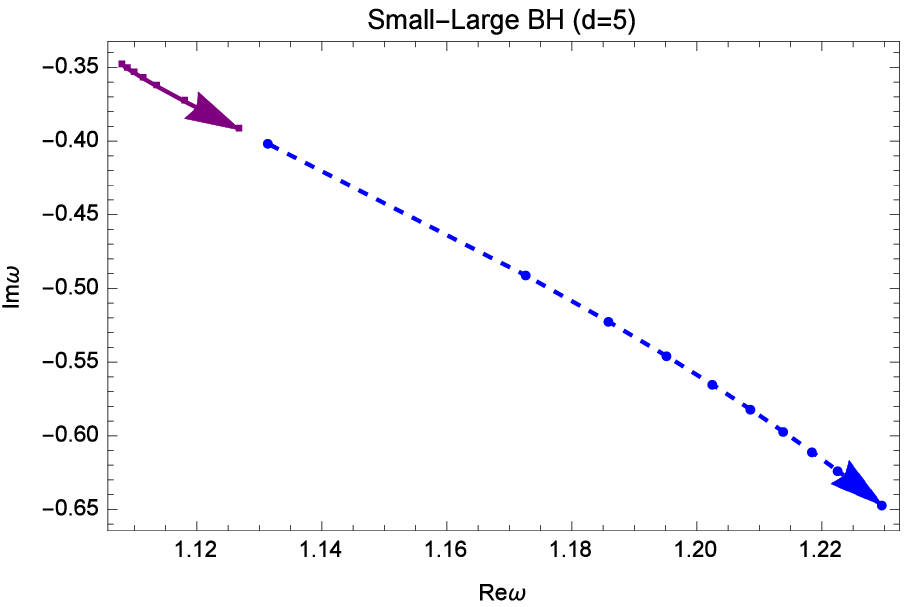} \>
				\includegraphics[width=4.5cm,height=3.cm]{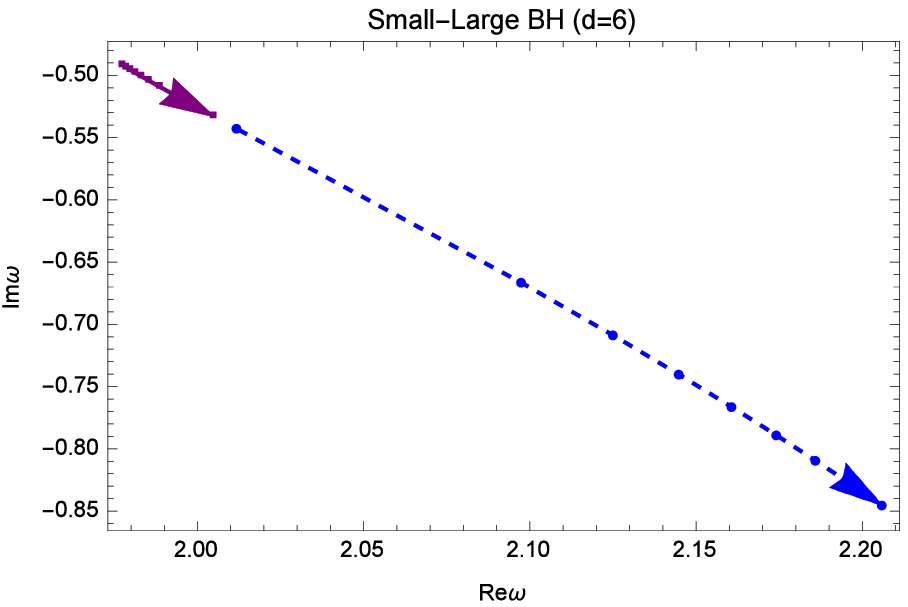}\> \includegraphics[width=4.5cm,height=3.cm]{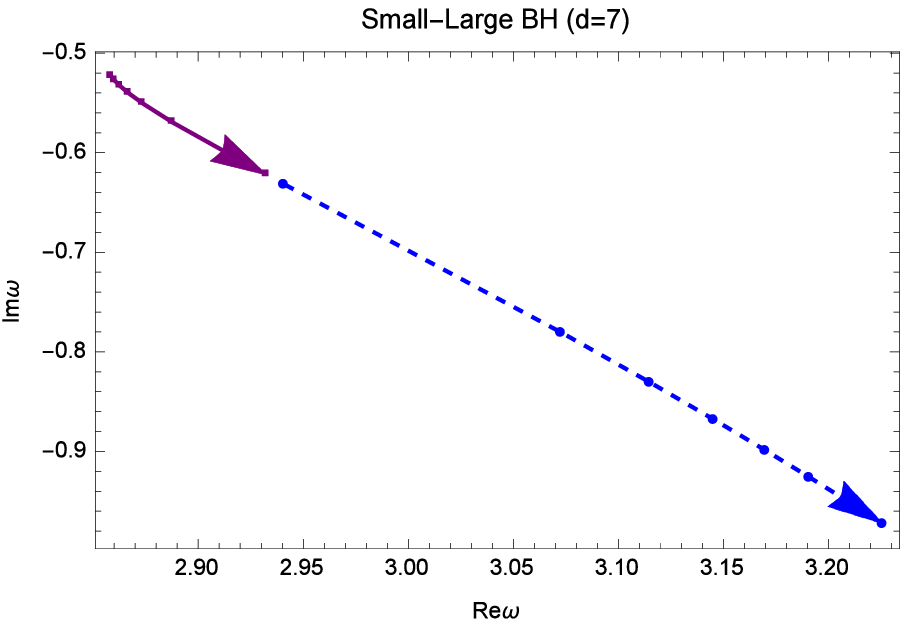} \\
				\includegraphics[width=4.5cm,height=3.cm]{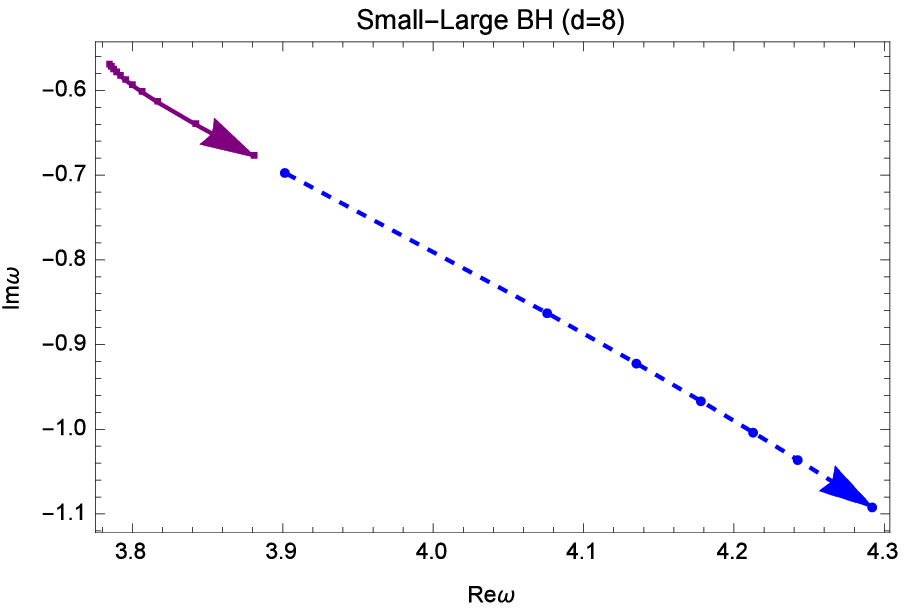}\> \includegraphics[width=4.5cm,height=3.cm]{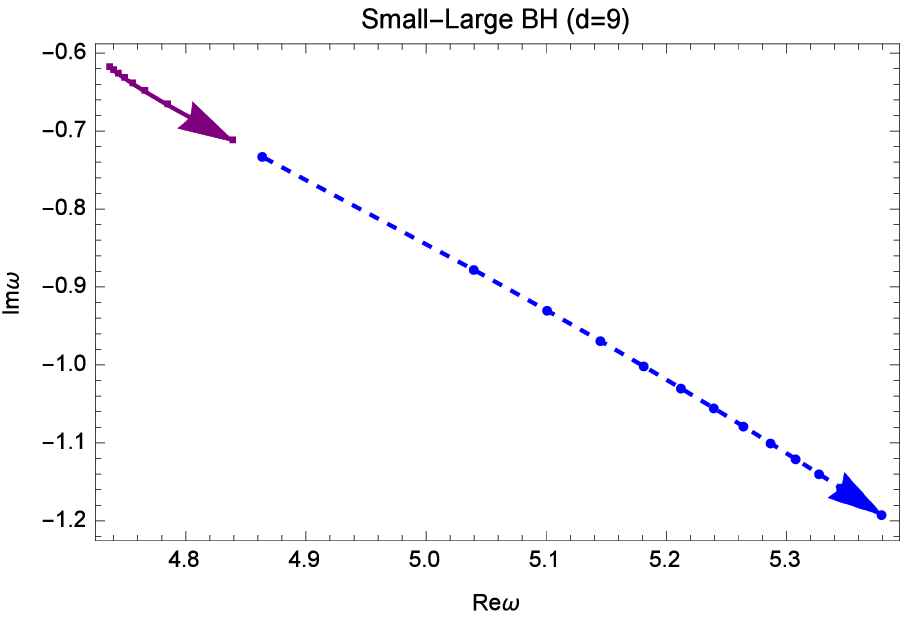} \>
				\includegraphics[width=4.5cm,height=3.cm]{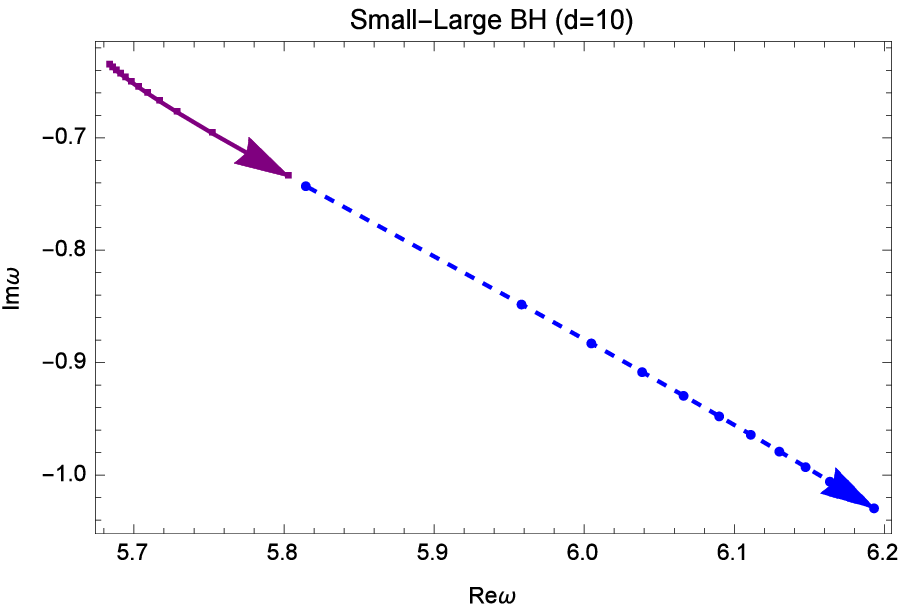}\> \includegraphics[width=4.5cm,height=3.cm]{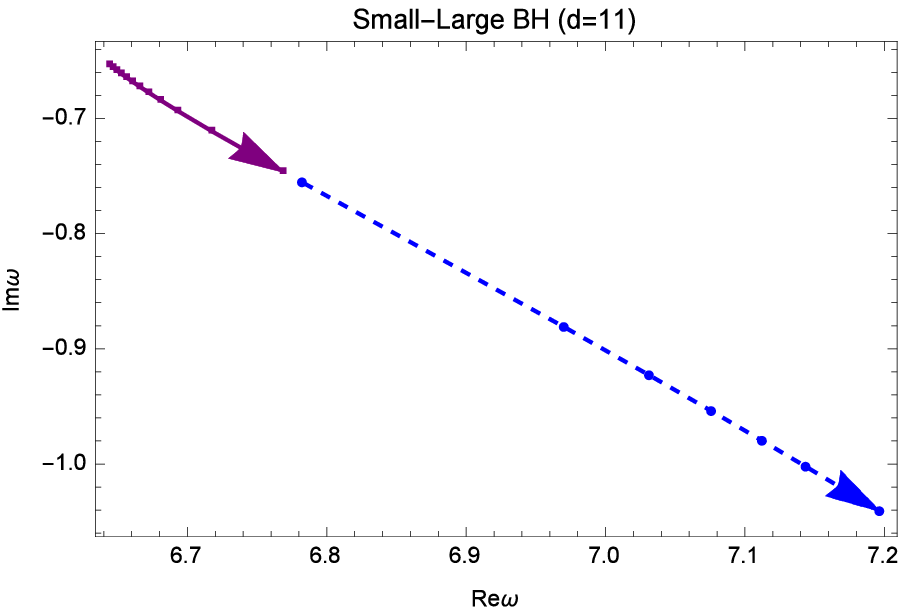} \\
			\end{tabbing}
			\vspace*{0cm} \caption{The behavior of the quasinormal modes for small (solid) and large (dashed) black holes in the isothermal second order  phase transition for a $4\leq d\leq 11$. Increase in the black hole size is shown by arrows.}\label{fig9}
		\end{minipage}
	\end{figure}

The quasinormal frequencies for small and large black hole phases are plotted in Figure \ref{fig9} for $4\leq d \leq 11$. For the isothermal phase transition at $T=T_c$, we can see that when phase transitions are realized, both the SBH and LBH QNMs have the same behavior as the horizon increases, which generalizes the result of  \cite{base} to higher dimensional spacetime.

\section{Conclusion}
In this work, we have investigated the thermodynamical behaviors of $d$-dimensional charged AdS black holes and their phase transitions.
 Emphasis has been put on the calculation  of the quasinormal modes frequencies of the massless scalar perturbation 
around small and large black holes. This study has been performed for isobaric process as well as isothermal one.
\\
For the former, below the critical pressure, we found out that for any dimension $d$ in the range $4\leq d\leq 11$, the QNMs can well be a dynamical tool to probe the black hole phase transition, since the slopes of the quasinormal frequency change between small and large black holes. As to the latter process,  for the phase transition below the critical point, we have shown that the spacetime dimension affects the behavior of QNMs. More precisely, we found that the change in QNM slopes occurs only for $d=4$ and $d=5$.
\\
Finally we have also found that, at the critical  isothermal and isobaric phase transitions, both small and large black holes quasinormal frequencies have  the same behavior, suggesting  that  quasinormal modes are not  appropriate to probe the black hole phase transition in the second order.
\newpage

\section*{Acknowledgements}
This work is supported in part by the Groupement de recherche international (GDRI): Physique de l'infiniment petit et de l'infiniment grand - P2IM. S.I would like to thank Prof. S. Mahapatra for providing a code to cross-check the calculations.


\begin{thebibliography}{99}

\bibitem{base} 
  Y.~Liu, D.~C.~Zou and B.~Wang,
  {\em Signature of the Van der Waals like small-large charged AdS black hole phase transition in quasinormal modes},
  JHEP {\bf 1409} (2014) 179,
  [arXiv:1405.2644].

\bibitem{vis}
C.~V.~Vishveswara, {\em Scattering of Gravitational Radiation by a Schwarzschild Black-hole}, Nature  {\bf 227} (1970) 936; {\em Stability of the Schwarzschild Metric}, Phys.\ Rev.\  {\bf D 1}  (1970)  2870. 

\bibitem{mx23} 
  H.~P.~Nollert,
  {\em TOPICAL REVIEW: Quasinormal modes: the characteristic `sound' of black holes and neutron stars},
  Class.\ Quant.\ Grav.\  {\bf 16} (1999) R159.

\bibitem{mx25} 
  E.~Berti, V.~Cardoso and A.~O.~Starinets,
  {\em Quasinormal modes of black holes and black branes},
  Class.\ Quant.\ Grav.\  {\bf 26} (2009) 163001,
  [arXiv:0905.2975].

\bibitem{mx1} 
  B.~P.~Abbott {\it et al.} [LIGO Scientific and Virgo Collaborations],
  {\em Observation of Gravitational Waves from a Binary Black Hole Merger},
  Phys.\ Rev.\ Lett.\  {\bf 116}  (2016)  061102,
  [arXiv:1602.03837].`

\bibitem{mx26} 
  B.~P.~Abbott {\it et al.} [LIGO Scientific Collaboration],
  {\em LIGO: The Laser interferometer gravitational-wave observatory},
  Rept.\ Prog.\ Phys.\  {\bf 72} (2009) 076901,
  [arXiv:0711.3041].

\bibitem{mx27} 
  T.~Accadia {\it et al.},
  {\em Status of the Virgo project},
  Class.\ Quant.\ Grav.\  {\bf 28} (2011) 114002.

  \bibitem{mx28} 
  K.~Somiya [KAGRA Collaboration],
  {\em Detector configuration of KAGRA-The Japanese cryogenic gravitational-wave detector},
  Class.\ Quant.\ Grav.\  {\bf 29} (2012) 124007,
  [arXiv:1111.7185].
  
  \bibitem{mx28b} 
  M.~Armano {\it et al.},
  {\em Sub-Femto- g Free Fall for Space-Based Gravitational Wave Observatories: LISA Pathfinder Results},
  Phys.\ Rev.\ Lett.\  {\bf 116} no. 23, (2016) 231101.
  
  \bibitem{cliff24}
  J.~M.~Maldacena,
  {\em The Large N limit of superconformal field theories and supergravity},
  Int.\ J.\ Theor.\ Phys.\  {\bf 38} (1999) 1113; 
  [Adv.\ Theor.\ Math.\ Phys.\  {\bf 2} (1998) 231],
  [hep-th/9711200].
  
  \bibitem{mx29} 
  V.~Cardoso, Ó.~J.~C.~Dias, G.~S.~Hartnett, L.~Lehner and J.~E.~Santos,
  {\em Holographic thermalization, quasinormal modes and superradiance in Kerr-AdS},
  JHEP {\bf 1404} (2014) 183, 
  [arXiv:1312.5323].
  
   \bibitem{mx30} 
   V.~Cardoso, R.~Konoplya and J.~P.~S.~Lemos,
   {\em Quasinormal frequencies of Schwarzschild black holes in anti-de Sitter space-times: A Complete study on the asymptotic behavior},
   Phys.\ Rev.\ D {\bf 68} (2003) 044024 ,
   [gr-qc/0305037].
   
   \bibitem{mx31} 
   C.~M.~Warnick,
   {\em On quasinormal modes of asymptotically anti-de Sitter black holes},
   Commun.\ Math.\ Phys.\  {\bf 333}  (2015) 959,
   [arXiv:1306.5760].
   
    \bibitem{mx30b} 
    R.~A.~Konoplya,
    {\em Decay of charged scalar field around a black hole: Quasinormal modes of R-N, R-N-AdS black hole},
    Phys.\ Rev.\ D {\bf 66} (2002) 084007, 
    [gr-qc/0207028].
    
    \bibitem{mx30b2} 
    R.~A.~Konoplya and A.~Zhidenko,
    {\em Stability of higher dimensional Reissner-Nordstrom-anti-de Sitter black holes},
    Phys.\ Rev.\ D {\bf 78} (2008) 104017, 
    [arXiv:0809.2048].
    
    \bibitem{mx31b} 
    R.~Li, H.~Zhang and J.~Zhao,
    {\em Time evolutions of scalar field perturbations in D-dimensional Reissner–Nordström Anti-de Sitter black holes},
    Phys.\ Lett.\ B {\bf 758} (2016) 359,  
    [arXiv:1604.01267].
   
  \bibitem{mx28b2} 
  R.~A.~Konoplya and A.~Zhidenko,
  {\em Quasinormal modes of black holes: From astrophysics to string theory},
  Rev.\ Mod.\ Phys.\  {\bf 83}, (2011) 793,
  [arXiv:1102.4014].
   
   
   \bibitem{lab1} 
  E.~Abdalla, R.~A.~Konoplya and A.~Zhidenko,
  {\em Perturbations of Schwarzschild black holes in laboratories},
  Class.\ Quant.\ Grav.\  {\bf 24}, 5901 (2007),
  [arXiv:0706.2489 [hep-th]].
  
  \bibitem{lab2} 
  R.~A.~Konoplya and A.~Zhidenko,
  {\em Stability of multidimensional black holes: Complete numerical analysis},
  Nucl.\ Phys.\ B {\bf 777}, 182 (2007),
  [hep-th/0703231 [hep-th]].
  
  
  \bibitem{lab3} 
  A.~Zhidenko,
  {\em Quasinormal modes of brane-localized standard model fields in Gauss-Bonnet theory},
  Phys.\ Rev.\ D {\bf 78}, 024007 (2008),
  [arXiv:0802.2262 [gr-qc]].
  
  \bibitem{lab4} 
  E.~Abdalla, C.~B.~M.~H.~Chirenti and A.~Saa,
  {\em Quasinormal mode characterization of evaporating mini black holes},
  JHEP {\bf 0710}, 086 (2007)
  doi:10.1088/1126-6708/2007/10/086
  [gr-qc/0703071].
        
  
\bibitem{01}
A.~Dabholkar, S.~Nampuri, {\em   Lectures on Quantum Black Holes},
Lect.Notes Phys. {\bf 851} (2012) 165.

\bibitem{02}
S.~J.~Rey, {\em String Theory on Thin Semiconductors: Holographic
Realization of Fermi Points and Surfaces},
Prog. Theor. Phys. Suppl. {\bf 177} (2009) 128.

\bibitem{04}
 H.~Witek, H.~Okawa, V.~Cardoso, L.~Gualtieri, C.~Herdeiro, M.~Shibata,
U.~Sperhake and M.~Zilhao, {\em Higher dimensional Numerical
Relativity: code comparison},   Phys. Rev. {\bf D 90} (2014) 084014, [arXiv:1406.2703].

\bibitem{Kastor:2009wy}
  D.~Kastor, S.~Ray and J.~Traschen,
  {\em Enthalpy and the Mechanics of AdS Black Holes},
  Class.\ Quant.\ Grav.\  {\bf 26} (2009) 195011
  [arXiv:0904.2765].

\bibitem{30} S.~Hawking and D.~N.~Page, {\em Thermodynamics of Black Holes
in Anti-de Sitter Space},   Commun. Math. Phys. {\bf 83} (1987) 577.

\bibitem{a1}
D.~Kastor, S.~Ray and J.~Traschen, {\em Enthalpy and the Mechanics of AdS
Black Holes}, Class. Quant. Grav. {\bf 261} (2009) 95011, [arXiv:0904.2765].

\bibitem{4} A.~Chamblin, R.~Emparan,  C.~Johnson and R.~Myers, {\em  Charged AdS black holes and catastrophic holography},
Phys. Rev. {\bf D 60} (1999) 064018.

\bibitem{5} A.~Chamblin, R.~Emparan, C.~Johnson and  R.~Myers,
{\em  Holography, thermodynamics, and fluctuations of charged AdS
black holes}, Phys. Rev. {\bf D 60}  (1999) 104026.

\bibitem{50} M.~Cvetic, G.~W.~Gibbons, D.~Kubiznak and C.~N.~Pope, {\em Black Hole Enthalpy and an
Entropy Inequality for the Thermodynamic Volume}, Phys. Rev. {\bf D
84} (2011) 024037, [arXiv:1012.2888].

\bibitem{6} 
B.~P.~Dolan, D.~Kastor, D.~Kubiznak, R.~B.~Mann and J.~Traschen, {\em Thermodynamic Volumes and Isoperimetric Inequalities for de Sitter Black Holes},
Phys. Rev. {\bf D 87}  no. 10, (2013) 104017, 
[arXiv:1301.5926].


\bibitem{KM} D.~Kubiznak and  R.~B.~Mann, {\em P-V criticality of charged AdS black holes}, JHEP
 {\bf 1207} (2012) 033.

\bibitem{chin1} C.~Song-Bai, L.~Xiao-Fang and L.~Chang-Qing, {\em $P$-$V$ Criticality of an AdS Black Hole in f(R) Gravity},
 Chin. Phys. Lett.,  {\bf 30} (2013) 060401.

\bibitem{our} A.~Belhaj, M.~Chabab, H.~El Moumni and M.~B.~Sedra, {\em
On Thermodynamics of AdS Black Holes in Arbitrary Dimensions}, Chin. Phys. Lett.
{\bf 29} (2012) 100401.

\bibitem{our1} A.~Belhaj, M.~Chabab, H.~El Moumni, L.~Medari and M.~B.~Sedra, {\em
The Thermodynamical Behaviors of Kerr-Newman AdS Black Holes}, Chin. Phys. Lett.
{\bf 30} (2013) 090402.

\bibitem{our2} A.~Belhaj, M.~Chabab, H.~El Moumni, K.~Masmar and M.~B.~Sedra, {\em
Critical Behaviors of 3D Black Holes with a Scalar Hair}, IJGMMP
{\bf12}  (2015) 1550017, [arXiv:hep-th/1306.2518.]

\bibitem{zzz1} 
P.~Prasia and V.~C.~Kuriakose,
\emph{quasinormal Modes and P-V Criticallity for scalar perturbations in a class of dRGT massive gravity around Black Holes},
Gen.\ Rel.\ Grav.\  {\bf 48}, no. 7,  (2016) 89
[arXiv:1606.01132].

\bibitem{Janiszewski}
S.~Janiszewski and M.~Kaminski, \emph{Quasinormal modes of magnetic and electric black branes versus far from equilibrium anisotropic fluids},
Phys.\ Rev.\  {\bf D 93} (2016)  025006, 
[arXiv:1508.06993].

\bibitem{Mahapatra} 
S.~Mahapatra,
\emph{Thermodynamics, Phase Transition and Quasinormal modes with Weyl corrections},
JHEP {\bf 1604}, (2016) 142 
[arXiv:1602.03007].

\bibitem{zhuWang}
J. M. Zhu, B. Wang, and E. Abdalla, \emph{Object picture of
	quasinormal ringing on the background of small Schwarzschild Anti-de
	Sitter black holes}, Phys. Rev.  {\bf D 63}, (2001) 124004,
[hep-th/0101133].

\bibitem{WangMolina}
B. Wang, C. Molina, and E. Abdalla, \emph{Evolving of a massless
	scalar field in Reissner-Nordstrom Anti-de Sitter spacetimes}, 
Phys. Rev. \  {\bf D 63} (2001) 084001,  [hep-th/0005143].  

\bibitem{Spallucci:2013osa}
  E.~Spallucci and A.~Smailagic,
  {\em Maxwell's equal area law for charged Anti-deSitter black holes},
  Phys.\ Lett.\ {\bf  B 723} (2013) 436, 
  [arXiv:1305.3379].


\bibitem{Spallucci:2013jja}
  E.~Spallucci and A.~Smailagic,
  {\em Maxwell's equal area law and the Hawking-Page phase transition},
  J.\ Grav.\  {\bf 2013} (2013) 525696, 
  [arXiv:1310.2186].


\bibitem{hasan}
 A.~Belhaj, M.~Chabab, H.~El moumni, K.~Masmar and M.~B.~Sedra,
  {\em Maxwell`s equal-area law for Gauss-Bonnet-Anti-de Sitter black holes},  Eur.\ Phys.\ J.\  {\bf C 75} (2015)  71, 
  [arXiv:1412.2162].

\bibitem{Zhao:2014eja}
  J.~X.~Zhao, M.~S.~Ma, L.~C.~Zhang, H.~H.~Zhao and R.~Zhao,
  {\em The equal area law of asymptotically AdS black holes in extended phase space},
  Astrophys.\ Space Sci.\  {\bf 352} (2014) 763.


\bibitem{Dolan}
  B.~P.~Dolan,
  {\em Vacuum energy and the latent heat of AdS-Kerr black holes},
  Phys.\ Rev.\  {\bf D 90} (2014)  084002,
  [arXiv:1407.4037].

\bibitem{Cliff}
  C.~V.~Johnson,
  {\em Holographic Heat Engines},
  Class.\ Quant.\ Grav.\  {\bf 31} (2014) 205002
  [arXiv:1404.5982].

\bibitem{14Dolan} S.~W.~Hawking and D.~N.~Page, {\em Thermodynamics of black holes in anti-de Sitter space}, Communications in Mathematical Physics, vol. {\bf 87}, no. 4 (1983) 577.
\bibitem{15Dolan}  N.~Altamirano, D.~Kubiznak, R.~B.~Mann and Z.~Sherkatghanad,
  {\em Thermodynamics of rotating black holes and black rings: phase transitions and thermodynamic volume},
  Galaxies {\bf 2} (2014) 89,
  [arXiv:1401.2586].


\bibitem{witten} E.~Witten, {\em Anti-de Sitter space, thermal phase transition, and confinement in gauge theories}.
Adv. Theor. Math. Phys.  {\bf 2} (1998)  505.
  


\end{thebibliography}
\end{document}